\documentclass[reprint,amsmath,amssymb,aps,pra,showpacs]{revtex4-1}

\usepackage{graphicx}
\usepackage{color}
\usepackage{epstopdf}
\usepackage{dcolumn}
\usepackage{bm}
\usepackage{bbm}
\usepackage{mathbbol}
\usepackage{dsfont}

\newcommand{\bwt}{\begin{widetext}}
\newcommand{\ewt}{\end{widetext}}
\newcommand{\bel}[1]{  \begin{equation} \label{#1}}
\newcommand{\eel}{\end{equation}}

\newcommand{\Traza}[1]{\mbox{Tr}\,\left\{#1\right\}}
\newcommand{\ket}[1]{\left| #1 \right>}
\newcommand{\bra}[1]{\left< #1 \right|}

\newcommand{\dd}{\mbox{d}}

\begin{document}

\title{Quantum Enhanced Energy Distribution for Information Heat Engines}
\author{J.M. Diaz de la Cruz$^{1}$ and M.A. Martin-Delgado$^{2}$}
\affiliation{$^1$Departmento de Fisica Aplicada e Ingenieria de Materiales, Universidad Politecnica, 28006 Madrid, Spain  \\ $^2$Departamento de Fisica Teorica I, Universidad Complutense, 28040 Madrid, Spain}

\preprint{APS/123-QED}

\begin{abstract}
A new scenario for energy distribution, security and shareability is presented that assumes the availability of quantum information heat engines and a thermal bath. It is based on the convertibility between entropy and work in the presence of a thermal reservoir. Our approach to the informational content of physical systems that are distributed between users is complementary to the conventional perspective of quantum communication. The latter places the value on the unpredictable content of the transmitted quantum states, while our interest focuses on their certainty. Some well-known results in quantum communication are reused in this context. Particularly, we describe a way to securely distribute quantum states to be used for unlocking energy from thermal sources. We also consider some multi-partite entangled and classically correlated states for a collaborative multi-user sharing of work extraction possibilities.  Besides, the relation between the communication and work extraction capabilities is analyzed and written as an equation.
\end{abstract}


\maketitle

\section{Introduction and antecedents}
The study of the relation between computational irreversibility and energy can be traced back as far as 1961 when Landauer\cite{Landauer1961} established that irreversible steps in a computation should dissipate an amount of heat directly related to the loss of information (an experimental demostration thereof has been published in 2012\cite{Berut2012}). On this line followed another important contribution by Bennett\cite{Bennett1973}. He claimed that any computation can be made reversible by the allocation of information in more registers (now known as ancillas), which can be deleted by the end of the process. Hamiltonian evolutions are reversible and, if they are time independent, do not dissipate any energy so that, under these circumstances, reversibility is directly related to energy conservation. Using quantum mechanics, in 1982 Feynman\cite{Feynman1982}, in 1984 Benioff\cite{Benioff1984} and in 1985 Deutsch\cite{Deutsch1985} proposed models for quantum computers with hamiltonian evolutions that minimized energy losses. A fairly up-to-date review of information and quantum computation can be found in \cite{Galindo2002}, and a comprehensive textbook in \cite{Nielsen2000}.

Another fruitful relation between energy and information began in the {\em Theory of Heat} published by J.C. Maxwell in 1871; he suggested that the introduction of an imaginary demon in a gas chamber could extract work from a single temperature gas volume. A further important contribution along this line is the Szilard's Engine\cite{Szilard1964} and countless studies later on paradoxes, contradictions or refinements of the said law. A lot of papers in the last years have elucidated the question from different perspectives, all of them confirming the validity of the second principle of thermodynamics \cite{Bennett1987,Zurek2003-1,Kim2011,Kish2012}.

Szilard engines and Maxwell demons analysis are often prone to misunderstandings. In our view they could be avoided with careful accounting for information and energy trade-off in the presence of feed-back systems\cite{Horowitz2011,Kish2012}. Generally, they use a {\em controller} which is governed by a number of bits (qubits in quantum environments). These bits are stored in some information container or {\em memory}. To change these bits it is necessary to reset them, thus some Landauer's work must be brought in and some entropy poured somewhere else. Then they must be replaced by new information coming from a {\em measurement} on the system, whereby some entropy is taken away. After that, some work can be obtained from the freshly measured bits. The process is then neutral in information and energy, unless entropy and work can be taken away in exchange of information. We would like to emphasize that the real fuel for these devices is information, as is explained in \cite{Diaz2014}for a magnetic quantum information heat engine, which is a new version of Szilard's engine devoid of some irrelevant details. An alternative to the measuring stage is periodically swapping high-entropy states in the engine with low-entropy ones supplied by feeding ancillas\cite{Uzdin2014,Diaz2014}.

It is now well understood that the availability of a thermal bath allows for a trade-off between information and work. The devices that carry out this conversion are known as {\em Information Heat Engines} \cite{Toyabe2010,Zurek2003-1,Scully2001,Scully2005,Rostovtsev2005, Sariyanni2005, Quan2006, Delrio2011, Kim2011, Plesch2013, Funo2013, Piechocinska2000, Zhou2010,Diaz2014,Lloyd1997,Vedral2016}. The information may be classical or quantum and the available energy could differ in both cases. The difference is defined by the {\em work deficit}\cite{Oppenheim2002} or {\em quantum discord} \cite{Park2013, Ollivier2002, Zurek2003-2}. Besides, the extractable work can be related to the relative entropy or Kullback-Leibler divergence\cite{Hasegawa2010, Vedral2002, Sagawa2012,Brandao2015}. Quantum statistical considerations with fermionic and bosonic multiparticle systems have also been analyzed elsewhere\cite{Lloyd1997,YuanjianZheng2015,Kim2011,Plesch2011,Diaz2014,
Jaramillo2015,Beau2016,Jeon2016}.

Other magnetic machines (classical and quantum) have been proposed for other different thermodynamic cycles, especially cooling systems\cite{Kosloff2010,Beretta2012,Gordon2009,Lloyd1997}.

The role of standard entropy in Information Heat Engines is relevant for average values of extracted work.  Generalizations to the so called {\em smooth} entropies have been made to include other fluctuations and probabilities of failure\cite{Dahlsten2011,Horodecki2011,Egloff2013}. This perspective is not considered in the present paper.

In this contribution a new scenario for communication and energy distribution is defined that allows the consideration of novel possibilities concerning the security and conditions of use for both the information and energy convertibility stored in a shared quantum state. 

Quantum information resources for communication have already been adapted to energy distribution\cite{Hotta2008pla, Hotta2009} and described using Ising chains\cite{Trevison2015}. They differ from our approach basically because they use an entangled vacuum state for a non-local hamiltonian which has to be shared by both the energy provider and consumer. On the other hand, our proposal includes the use of Quantum Information Heat Engines (henceforth QIHE) and the availability of a thermal bath at non-zero temperature.  

In section \ref{sec::QIHE} we provide a short description of a basic model for a QIHE, to which further ideas will be referred along the rest of the paper.  Section \ref{sec::divertimento} analyzes the transmission of physical systems, that will be referred to as {\em messengers}, whose entropy may be increased at a remote station in order to extract work from thermal baths; several protocols with different functionalities  are presented. For example, by using shared entangled states, qubits sent from an emitter $A$ to a receiver $B$ are completely depolarized\footnote{In this work, a completely depolarized system is equivalent to a system with maximum entropy. Taking into account that Information Heat Engines extract work by increasing the entropy of the fueling system, a completely depolarized state yields zero work.}, thus useless for any illegitimate user that could intercept them; other multipartite systems render other interesting possibilities concerning the conditions to access not completely depolarized systems.  Sections \ref{sec::limitation} and \ref{sec::simultaneous} analyze the transmission of physical systems in a given quantum state from two different perspectives which prove to be complementary. One of them is quantum communication, where new milestones\cite{Xiao2012,Herbst2015} are frequently announced, and the other is the possibility to fuel a QIHE under suitable restrictions. A relation between the communication and work extraction capabilities of sources of physical systems is sought and an equation is found. 
Section \ref{sec::simultaneous} specifically focuses on proving the possibility of reaching the Holevo bound for communication performance and simultaneously using the messenger system to fuel a QIHE, whereas in Section \ref{sec::limitation} both functions are supposed to be used exclusively.
Section \ref{sec::results} summarizes the conclusions.

\section{\label{sec::QIHE} Short description of a Quantum Information Heat Engine}  

\begin{figure}[h]
\begin{center}\includegraphics[width=0.4\textwidth]{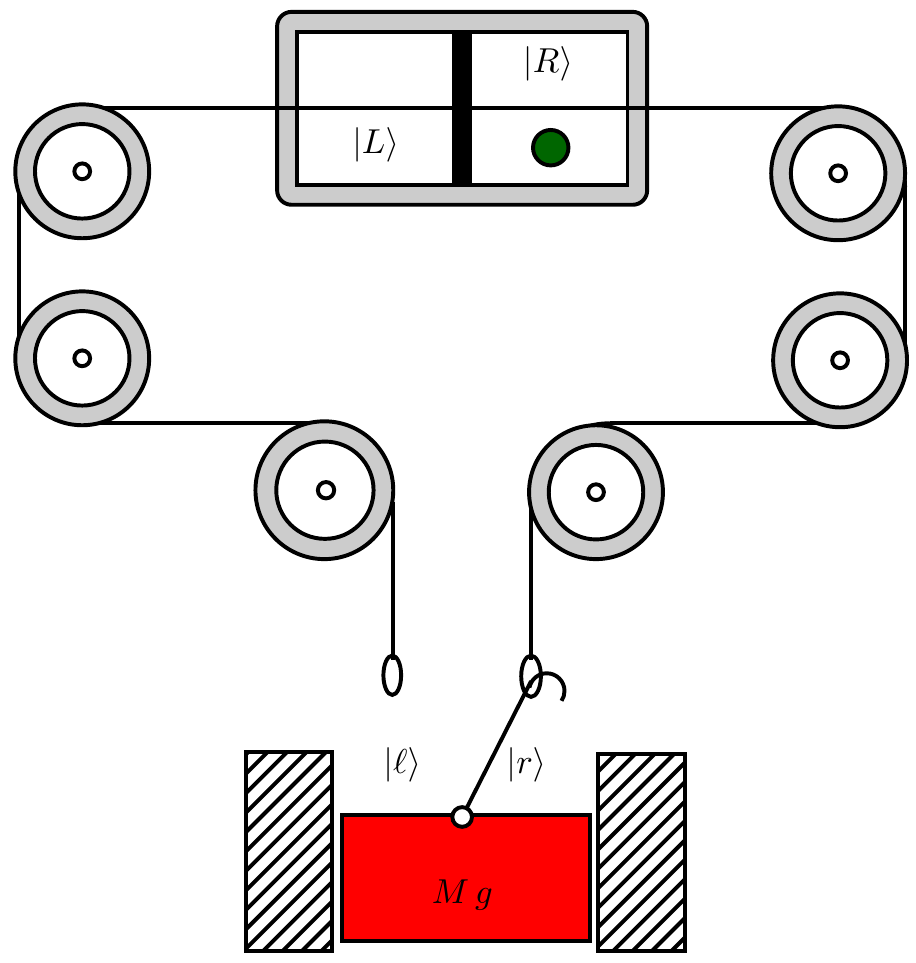}\end{center}
\caption{One particle Szilard cylinder engine; a measurement is carried out to determine whether the particle is in the $\ket{L}$ or $\ket{R}$ state and the result is stored in the $\ket{\ell}$ or $\ket{r}$ state of an indicator which, in turn, determines whether the mechanical plunder is attached to the left or right end of the cable depicted in the figure. The work stored in the potential energy of the plunder after full expansion of the barrier in the cylinder is $k_B\,(\ln 2)\,T$, and is exactly equal to the Landauer's work that would be needed to reset the indicator before the beginning of the next cycle. If the system works as a QIHE, instead of this energy, a fresh entropy-free indicator bit is supplied. }
\label{fig::szilard}
\end{figure}

Information Heat Engines are devices which cyclically convert thermal energy into useful work at the expense of degrading information. They are periodically fueled with auxiliary systems, whose entropy is increased, although their energy is conserved. The power delivered by the engine is drawn from a single thermal reservoir. This is in contrast with other cyclic engines that decrease the internal energy of the systems that are supplied as fuel. It should also be remarked that they comply with all three principles of thermodynamics. In the following we assume that the engine is a physical system with a local hamiltonian and no energy is exchanged with the fueling system. If the evolution is divided into intervals with either hamiltonian evolution or reversible thermal equilibrium with a bath at temperature $T$, the work after a complete cycle can be evaluated by
\bel{workcu}
W_{\rm cycle} \,=\,-\, \oint_{\rm cycle} \dd\left(\Traza{\rho H}\right) + \int_{\rm te} \,k_B\,(\ln 2)\,T\,\dd S\,,
\eel 
where the first integral extends for the whole cycle and vanishes on account of its periodic domain. The second integral is defined only for the intervals when the system is at thermal equilibrium with the bath. Besides, entropy is expressed in bits; $k_B$ is Boltzman's constant, $\rho$ is the density matrix of the system and $H$ represents the hamiltonian. The result is 
\bel{workc}
W_{\rm cycle} \,= \,k_B\,(\ln 2)\,T\,\Delta S\,,
\eel 
which shows that work can be obtained only if entropy is reset to a lower value at some point in the cycle. Entropy extraction can be accomplished in a number of ways. The simplest one is swapping the quantum state of the system with that of an ancilla with an initially lower value of entropy. More frequently, entropy is extracted through feedback control. In this case, the system is driven to a lower entropy state by first measuring and then triggering the suitable hamiltonian to make the system evolve to the desired state.  As it is thoroughly analyzed in \cite{Diaz2014}, a measurement involves two systems: the measurand and the indicator, which must start in a well defined state. After the measurement the indicator ends up with some entropy, on account of the unpredictability of the outcome. In contrast, the system is brought to a less entropic state by virtue of the tailored hamiltonian evolution triggered according to the outcome of the measurement. It is after the measuring stage that entropy has to be removed by resetting the indicator; the Landauer's work is precisely the product of this entropy and the temperature. We assume that the states of the indicator have negligible energy differences in order to isolate its purely informational role.

\begin{figure}[t]
\begin{center}\includegraphics[width=0.4\textwidth]{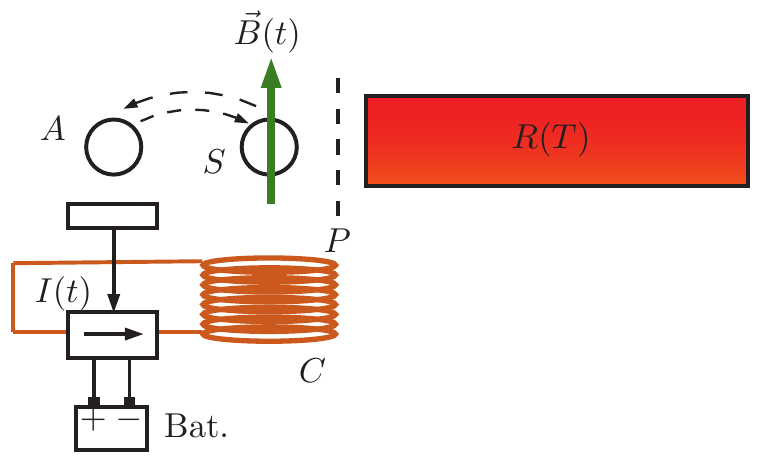}\end{center}
\caption{Magnetic Quantum Information Heat Engine; a spin-$\frac 1 2$ particle $S$ lies in the magnetic field generated by an electrical current $I(t)$ circulating through a coil $C$. An ancilla $A$ is used to measure the $z$ component of the magnetic moment of $S$; the result determines the value of the magnetic field generated by $I(t)$. Then the field is gradually turned off, while $S$ keeps in thermal equilibrium with a reservoir $R(T)$ at temperature $T$. At the end of the process an energy $W$ is stored at the electrical battery $B$; its value is $W=k_B\,(\ln 2)\,T$, on account of the increase of entropy of $S$, which enters the isothermal process with zero entropy and exits completely depolarized. Again, this energy is equal to the Landauer's energy needed to reset the ancilla for the next cycle. If the system works as a QIHE, instead of this energy, a fresh entropy-free ancilla bit is supplied . }
\label{fig::MQIHE}
\end{figure}

In a single particle cylinder Szilard engine, as shown in Fig.\ref{fig::szilard}, measurement of which-side is performed and isothermal expansion against a wall in the suitable direction follows. A barrier is then repositioned at the middle of the cylinder. The entropy increase between the moment following the measurement and immediately after the reinsertion of the barrier is exactly one bit, corresponding to the loss of information about which-side of the cylinder the particle is in. Correspondingly, the extracted work per cycle $W_{\rm cycle}^{\rm Sz}$ is
\bel{workcsz}
W_{\rm cycle}^{\rm Sz} \,= \,k_B\,(\ln 2)\,T\,.
\eel  

In a magnetic quantum information heat engine\cite{Diaz2014}, as shown in Fig.\ref{fig::MQIHE}, measurement of the $z$-component of the magnetic moment of a spin-$\frac 1 2$ particle is followed by the onset of a suitable magnetic field which is later gradually removed in a stage at thermal equilibrium with a thermal reservoir. Exactly as in the one particle Szilard cylinder, the entropy increase between the moment after the measurement and the restart of the process is one bit and Eq.\ref{workcsz} holds. 


 The resource theory behind energy extraction\cite{Brandao2013,HorodeckiOppenheim2013,Sagawa2009,Sagawa2012} establishes that the obtainable work is related to the free energy of the state that describes the ancilla; when the energy differences are negligible, the important magnitude is the entropy. These results are generalized to develop a general formalism to calculate the maximum
efficiency of work extraction from an arbitrary quantum system in \cite{DeLiberato2011}. It is then straightforward to envisage a system for distribution of {\em purity}. Basic quantum cryptographic procedures can be implemented  in order to lock it for all but a legitimate participant and render it useless for anyone else. Furthermore, we analyze the possibilities furnished by some particular entangled states to enforce a collaborative procedure for unlocking the {\em purity} of the ancilla state.


\section{\label{sec::divertimento} Scenario for energy distribution through quantum information in multipartite systems  }

In the previous sections it has been established that a physical system whose quantum state is not completely depolarized can be transformed into another state with greater entropy and obtain work in the process, provided the availability of a thermal bath and an Information Heat Engine, like a Szilard or a magnetic information engine as described in section \ref{sec::QIHE}. 

\begin{figure*}[t]\begin{center}
\includegraphics[width=0.75\textwidth]{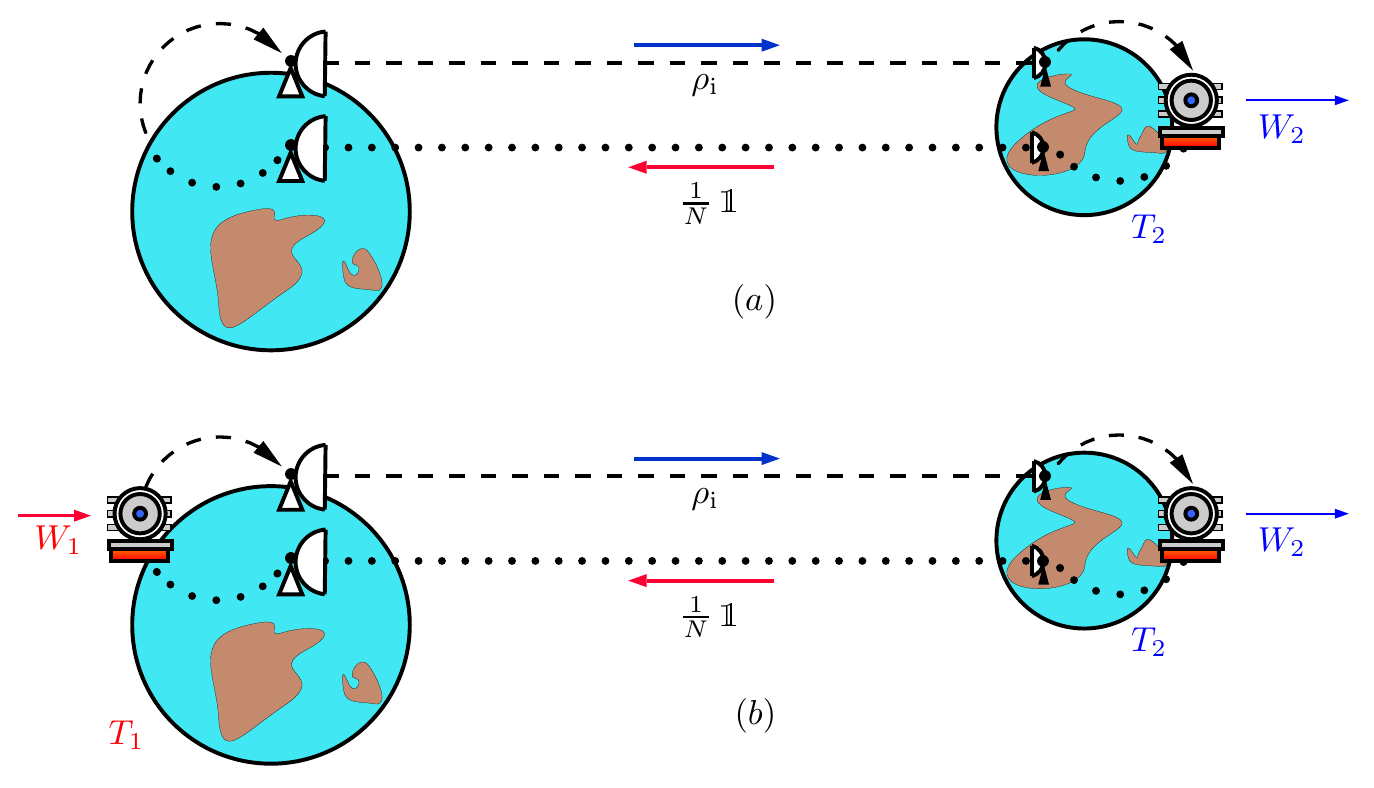}\end{center}
\caption{Part $(a)$ represents a way of supplying energy to a remote station at temperature $T_2$, consisting of sending physical systems in a quantum state $\rho_i$ whose entropy increases after fueling a QIHE and return as completely unpolarized states. As an example, photons can be sent to the remote station under a particular polarization and return completely unpolarized. The extracted work is $W_2\,=\,k_B\,(\ln 2)\,T_2$. In $(b)$ a remote Carnot cycle is depicted, where entropy is interchanged between the primary station at temperature $T_1$, where entropy is pumped out of the traveling physical system at a work cost $W_1\,=\,k_B\,(\ln 2)\,T_1$ , and the remote station. In this scenario, the net work gain per transmitted qubit is $W=W_2-W_1= \,k_B\,(\ln 2)\,(\,T_2-T_1)$ and, taking into account that the heat taken from $T_2$ is $Q_2\,=\,k_B\,(\ln 2)\,T_2$, it follows that $W=\,Q_2\,(\,1-\frac{T_1}{T_2})$ which is the standard result for Carnot cycles.}
\label{fig::telecarnot}
\end{figure*}

Now we set up a new scenario for energy distribution: a power information plant $P$ can do the reverse process and polarize a {\em messenger} physical system $M$ by supplying electrical (or mechanical or of any other kind) work (see Fig.\ref{fig::telecarnot} $(a)$). If the system $M$ (a photon polarization, for instance) is sent to other users $A,B,\ldots$ (may be in remote places) they can do the reverse operation and obtain work by depolarizing the system, provided they know the state of $M$. Note that if the reservoirs are at different temperatures this is equivalent to a (remote) Carnot cycle (see Fig.\ref{fig::telecarnot} $(b)$).

The next important issues about this scenario are the security and shareability. As a first scenario, one could send some messenger qubits from the power plant $P$ to $A$, so that $A$ can convert them into work. In order for $A$ to be able to obtain energy from the qubits, $A$ must know something about the state $\rho$ of said qubits. If all the qubits were in the same state, this could be learned by someone else (eavesdropper $E$) who could use them to draw some work from an available thermal reservoir. One possible solution that would avoid this reward for $E$, as we know from quantum security communication protocols, is to share a set of entangled qubits; then sending a qubit from $P$ would make $A$ obtain two fresh convertible bits (superdense coding). Thus, the same protocols that serve the purpose of protecting information can be used to secure that no unauthorized user converts the qubits into work. In short, we can implement a cryptographic protocol that renders the $S$ qubits completely unpolarized (useless) for any eavesdropper while carrying two convertible bits for legitimate users. 

As an example, $A$ and $B$ may agree on sharing a set of Bell pairs. Let $\ket{\Psi}_{\rm AB}$ be one of them. As it is widely known, the partial density matrices describing the $A$ and $B$ parts of $\ket{\Psi}_{\rm AB}$ are completely depolarized, being:
\bel{partab}
\rho_{\rm A} = \rho_{\rm B} = \frac 1 2 \left(
\begin{array}{cc}
1&0\\0&1
\end{array}
\right)\,,
\eel 
or, in other words, no information is available in the local subsystems. However, when $A$ sends its qubit to $B$ through a quantum channel, $B$ can reconstruct the bipartite initial Bell state, which is a pure state and can be used to extract a work $W_{\rm BS}$ given by
\bel{WBS}
W_{\rm BS} \,=\, 2 \,k_B \,(\ln 2)\, T_{B}\,.
\eel
It is important to note that interception of the qubit being sent from $A$ to $B$ is useless for any illegitimate user, on account of its total depolarization; only the two qubits together can be fruitful. Besides, it should also be remarked that the whole entropy generated in resetting the Bell pair is extracted from the thermal bath at $B$, so that nothing is lost. As a further remark, if thermal baths at different temperatures $T_A,T_B$ are used as sources or sinks of entropy in $A,B$, the system as a whole functions as a remote direct Carnot cycle if $T_A<T_B$ or a reverse one when $T_B>T_A$. 

A classical version of this protocol can be devised where a set of completely correlated random bits would be used instead of the Bell states. However, completely correlated pairs of random bits carry one bit of entropy and consequently, the extractable work $W_{\rm CB}$ would be
\bel{WCB}
W_{\rm BS} \,=\, k_B \,(\ln 2)\, T_{B}\,,
\eel
which is only half of the one given by Eq.\ref{WBS}. This reduction can be related to the superdense coding feature of quantum communication, i.e. one qubit may be used to communicate two classical bits.

\begin{figure}[t]
\begin{center}\includegraphics[width=0.35\textwidth]{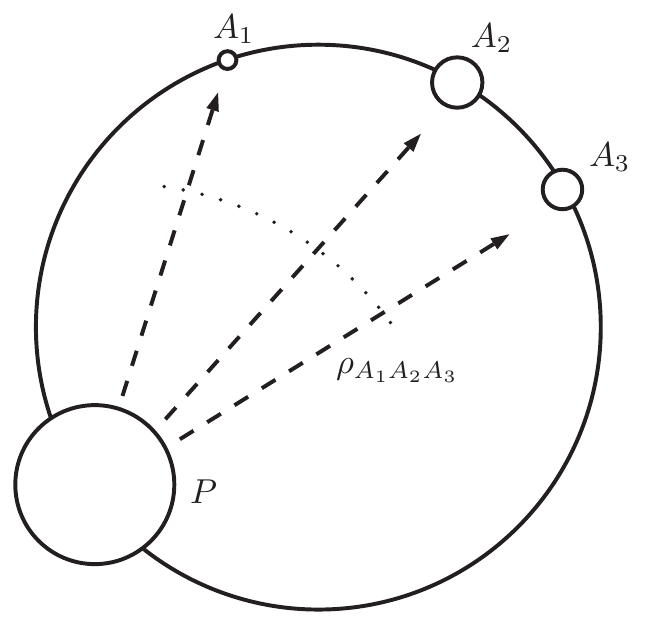}\end{center}
\caption{The source $P$ sends $\rho_{A_1A_2A_3}$  states to remote systems ${A_1,A_2,A_3}$.}
\label{fig::ring}
\end{figure}

There are further possibilities: one can use multipartite states $\rho_{A_1,\ldots,A_n}$ (see Fig. \ref{fig::ring}) that allow energy extraction under one of the following two situations:
\begin{itemize}
\item[i.-] any user $A_i$ can independently make that the other ones ($A_{j\neq i}$) access the energy of 1 bit in their part of $\rho_{A_1,\ldots,A_n}$. This possibility can be implemented by using generalized GHZ states, also known as cat states. An $n$-partite GHZ (Greenberger-Horne-Zeilinger) state is a pure state: 
\bel{GHZn}
\ket{\mbox{GHZ}_n} :=2^{-1/2}\left( \ket{0}^{\otimes n}+\ket{1}^{\otimes n}\right)
\,,\eel
with two remarkable properties:  1) any local measurement destroys the entanglement and 2) the partial states are completely unpolarized (useless for energy conversion). User $A_i$ can measure its qubit and then classically broadcast the result to all other users who can subsequently convert their $n-1$ bits into work.
\item[ii.-] only if all users agree, one of them can convert the energy. For this purpose we may use a classically correlated mixed state:
\bel{EPn}
\rho_{EP}:= \frac{1}{2^{n}}\,\sum_{i_1,\ldots,i_n}(1+(-1)^{i_1+\ldots+i_n})\,\sigma_{{i_1,\ldots,i_n}}
\,,
\eel
where $i_k$ may be $0$ or $1$ and, for notational simplicity, we define
\bel{defn}
\sigma_{{i_1,\ldots,i_n}}:=\ket{i_1,\ldots,i_n}\bra{i_1,\ldots,i_n} \,.
\eel

According to Eq.\ref{EPn}, $\rho_{EP}$ is a mixture of all even parity pure states. It can only furnish one bit of information upon knowing all but one bits measured on the standard basis. We will now prove this. For that purpose, we will show that any operation on $n-2$ qubits, say $3,\ldots,n$, gives no information about qubit $1$. Let $K_j$ be the Krauss operators; then, ignoring a possible normalization factor, the state of the system after it is:
\bel{EPp}
\rho'= \sum_j K_j \rho_{EP} K_j^\dagger \,,
\eel
where $K_j := \mathbbm{1}_2\otimes\mathbbm{1}_2\otimes k_j$, what accounts for the fact that the operation is on the last $n-2$ qubits. Expanding $\rho_{EP}$ we have
\bel{EPs}
\rho'=  \sum_j \sum_{i_1,\ldots,i_n\in {E}} k_j \sigma_{i_3,\ldots,i_n}k_j^\dagger \otimes \sigma_{i_1,i_2}\,.
\eel
Next, we separate the $i_3,\ldots,i_n\in {E}$ (set of even parity $n-2$-uples of bits)  and $i_3,\ldots,i_n\in {O}$ (set of odd parity $n-2$-uples of bits) contributions so that 
\begin{eqnarray*}\hspace{-5mm}
\rho'=& & \hspace{-6mm}\sum_{i_3,\ldots,i_n\in {E},j} \hspace{-5mm} k_j \sigma_{i_3,\ldots,i_n} k_j^\dagger \otimes (\sigma_{0,0}+\sigma_{1,1})
\\ &+& \hspace{-6mm} \sum_{i_1,\ldots,i_n\in {O},j} \hspace{-5mm} k_j \sigma_{i_3,\ldots,i_n} k_j^\dagger \otimes (\sigma_{0,1}+\sigma_{1,0}).
\end{eqnarray*}
Subsequently, we trace over all but the first and second qubits to obtain $\rho'_{12}$; let the constants $C_E,C_O$ be defined by:
\bel{Cs}
C_{E/O}:=\sum_{\substack{i_3,\ldots,i_n\in {E/O}\\ \ell_3,\ldots,\ell_n},j} \hspace{-5mm}|\bra{\ell_3,\ldots,\ell_n}k_j \ket{i_3,\ldots,i_n}|^2,
\eel
so that the partial trace results
\bel{ty}
\rho'_{12} = C_{E} (\sigma_{0,0}+\sigma_{1,1})+ C_{O} (\sigma_{0,1}+\sigma_{1,0}),
\eel
and finally, tracing over the second qubit we obtain:
\bel{typ}
\rho'_{1} = C_{E} (\sigma_{0}+\sigma_{1})+ C_{O} (\sigma_{1}+\sigma_{0}) = (C_{E}+C_{O})\,\mathbbm{1}_1,
\eel
which shows that no information is gained about qubit $1$, as it had been previously anticipated. 
\end{itemize}

\section{\label{sec::limitation} Mutual limitation between communication and energy extraction}

According to the previous paragraphs, posting messenger quantum states from $P$ to $A$ may serve one or two of the following purposes: sending qubits to be converted into energy and communicating information. Both features are mutually exclusive functions whose limiting relation is next obtained.

Holevo\cite{Holevo1998} in 1998, and Schumacher et al.\cite{Schumacher1997} in 1997,  established that given an alphabet consisting of 1) a set of ${\cal N}$ quantum mixed states (also known as letters) described by their density matrices $\{\rho_a,\,a = 1\mbox{ through }{\cal N}\}$ and 2) a set of probabilities $\{p_a,\,a = 1\mbox{ through }{\cal N}\}$ according to which the states have to be used, one can devise a communication protocol to convey $\chi$ bits (in average for sufficiently long messages) of classical information per state in the alphabet, where $\chi$ is defined as the {\em Holevo information} of the alphabet:\bel{holdef}
\chi:=S(\rho_B)-\left<S_a\right>
\,,\eel  
where 
\bel{S1}
\rho_B:=\sum_{a=1}^{\cal N} p_a \rho_a
\eel
is the mixed state that exits the emitter $A$ according to the information of the receiver $B$ and 
\bel{Saa}
<S_a>:=\sum_{a=1}^{\cal N} p_a S(\rho_a)\,.
\eel
It is understood that the communication protocol assigns a sequence of states in the alphabet, or {\em codeword} to every message that can be transmitted. 

We are now going to use this theorem\cite{Wilde2013} to state a new interesting result.

Let us now describe a bipartite scenario, with $A$ (Alice) and $B$ (Bob). A messenger physical system $M$ can be emitted from $A$ to $B$. The states of $M$ belong to a finite set $\rho\,\in\,\{\rho_B,\ldots,\rho_{\cal N}\}$. Besides, the availability of the states of $M$ follows a probability distribution $\{p_1,\ldots,p_{\cal N}\}$ which determines the frequencies with which the states of $M$ have to be used. Thus, the states $\{\rho_a,\,a = 1\mbox{ through }{\cal N}\}$ with probabilities $\{p_a,\,a = 1\mbox{ through }{\cal N}\}$ make the letters of an alphabet. If Alice and Bob agree on a certain code, the ordering of the states sent from $A$ to $B$ may be used to communicate messages, according to the Holevo information of the alphabet. Besides, not completely depolarized systems can also be used to fuel a QIHE, provided that there is a thermal bath at $B$.  

\begin{figure}[t]\begin{center}
\includegraphics[width=0.45\textwidth]{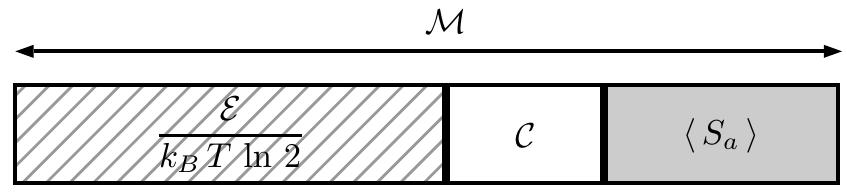}\end{center}
\caption{When some source of quantum states is used to convey classical information, it may also serve as an ancilla to extract work from a thermal source. Eq.\ref{ml} sets a mutual limitation for both functions. This figure represents a distribution of the functionality of a source $\{\rho_a,p_a\}$ per transmitted letter between energy conversion ${\cal E}$ (hatched), communication ${\cal C}$ (white) and a useless part $\left<S_a\right>$ (gray), showing that they add up to ${\cal M}$.}
\label{fig::conjuntos}
\end{figure}

According to Eq.\ref{workc}, the average work ($\cal E$) obtainable from each letter reads 
\bel{cale}{\cal E}=k_B T \,(\ln 2)\,\left({\cal M}-S(\rho_B)\right)\,,
\eel where
\bel{calM}{\cal M}:=\,\log_2\,d
\eel
and $d$ is the dimension of the Hilbert space for the physical system $M$; ${\cal M}$ is then the equivalent number of qubits per letter in the alphabet\footnote{Note that ${\cal M}$ is  not generally an integer number.}. Accordingly, substituting for $S(\rho)$ in Eq.\ref{holdef} yields
\bel{ml}
\frac{\cal E}{k_B T \ln 2}+{\cal C}+\left<S_a\right> = {\cal M}\,,
\eel
where ${\cal C}$ is the average information per emitted letter that one can communicate using the alphabet $\{\rho_a,p_a\}$ (which, according to the previous paragraph, is equal to $\chi$). Eq.\ref{ml} states the mutual limitation between communication and energy as is graphically represented in Fig.\ref{fig::conjuntos}.  

\section{\label{sec::simultaneous}Simultaneous supply of information and work extraction capability}

The derivation of Eq.\ref{ml} in the previous section was based on independent results for communication and thermodynamics scenarios. It proves that there is a relation between two different functions of the states of a system $M$, when they are used exclusively, that is for either sending messages or fueling a QIHE. Next, we show that both functions can be obtained simultaneously for the same system with a mutual limitation of Eq.\ref{ml}. For that, we refer to the decoding of information described in \cite{Schumacher1997}. Briefly stated, the results put forward in that paper which are relevant to this work are:

Given any positive real numbers $\epsilon,\delta$ and a set of states of a physical system $M$ with Hilbert space ${\cal H}$, $\{\rho_a,\,a = 1\mbox{ through }{\cal N}\}$ with probabilities $\{p_a,\,a = 1\mbox{ through }{\cal N}\}$, then for a sufficiently large integer $L$, 
\begin{enumerate}
\item there is a typical subspace $\Lambda_{L,\epsilon,\delta}\,\subset\,{\cal H}^{\otimes\,L}$, whose orthogonal projector is $\Pi_{L,\epsilon,\delta}$ that verifies
\bel{prob}
\Traza{\Pi_{L,\epsilon,\delta}\,\rho_L}>1-\epsilon,
\eel  
where $\rho_L:=\rho_B^{\otimes\,L}$, and 
\bel{defru}
\rho_B:=\sum_{a=1}^{\cal N} p_a \rho_a
\eel
in the following we drop the subscripts $L,\epsilon,\delta$ and simply write $\Lambda$ to denote the subspace $\Lambda_{L,\epsilon,\delta}$.
\item the dimension $d_\Lambda$ of $\Lambda$ is bounded by 
\bel{dimbound}
d_\Lambda :=\,\dim\,(\Lambda)\,\leq\,2^{L(\delta + S(\rho_B))} 
\eel
\item there is a communication protocol between Alice and Bob, whose information is coded\footnote{In \cite{Schumacher1997} the codes are constructed by choosing a number of codewords independently, according to the {\em a priori} string
probability for each codeword. The choice is supposed to be random and the results concerning the probability of error are averaged over the different possibilities.} in the order in which Alice arranges the letters $\rho_a$ before sending them to Bob, where the frequency of letter $\rho_a$ is $p_a$, that achieves $\chi-5\delta$ bits per letter with a probability of error $P_E\leq 10\epsilon$. When Bob receives the codewords sent by Alice, he performs a POVM (POM in the reference article\cite{Schumacher1997}), defined by a collection of {\em effects}  $E_j=\ket{\mu_j}\,\bra{\mu_j}$, where all of the kets $\ket{\mu_j}$ belong to the typical subspace $\Lambda$. All the ensuing process of decoding relays only on the result of this POVM.
\end{enumerate}

Our next purpose is to use this result to closely factorize the state $\rho_L=\rho_B^{\otimes\,L}$ into two parts: one which carries the information (and consequently should contain most of the entropy) and other which can fuel a QIHE (which, inversely, should have low entropy). As a first attempt we would want to factorize the $d_L:=2^{L {\cal M}}$-dimensional Hilbert space into a $d_\Lambda$ dimensional one and another which should accordingly have dimension $d_L/d_\Lambda$. This can  only happen when $d_L/d_\Lambda$ is a positive integer number, which is not the case in a general situation. 

In order to guarantee the possibility of factorization in the general case, we enlarge the Hilbert space ${\cal H}_L:={\cal H}^{\otimes\,L}$ by a tensor product with a $d_\Lambda$-dimensional ancillary physical system $D$ upon reception of the codeword by $B$. Let $\ket{0}_D$ be the initial state of the ancillary system $D$ and denote by ${\cal H}_D$ the Hilbert space of $D$. 

The enlarged codewords live now in the $d_L \times  d_\Lambda$-dimensional space defined by the tensor product ${\cal H}_L \,\otimes\,{\cal H}_D$. 
\begin{figure}[t]\begin{center}
\includegraphics[width=0.5\textwidth]{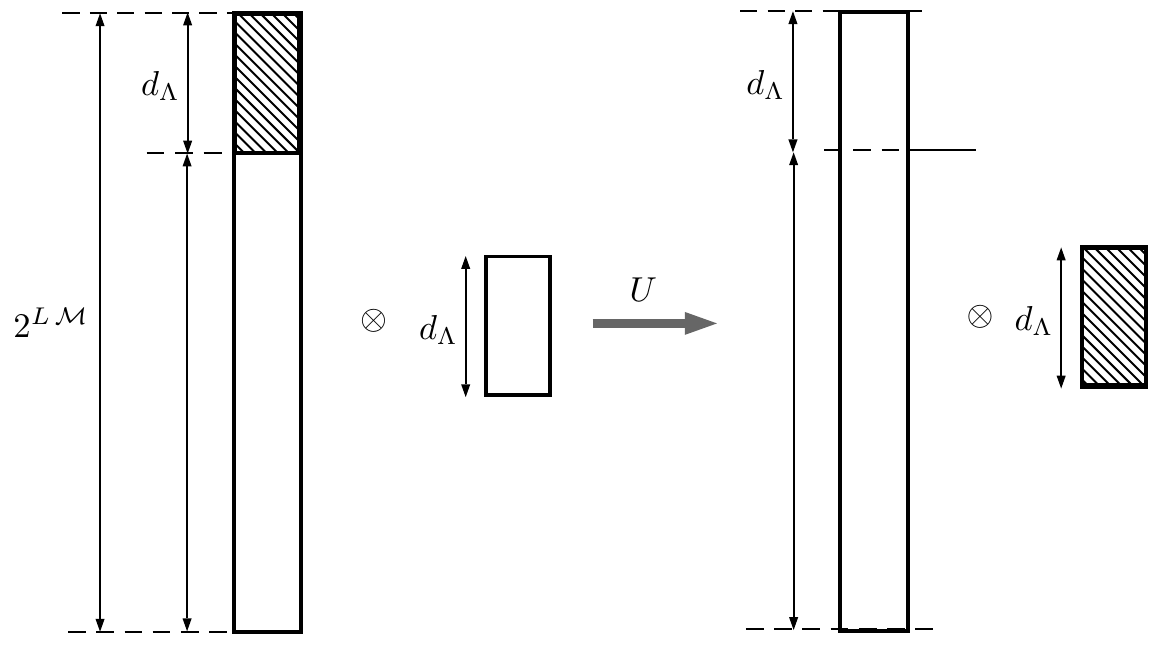}\end{center}
\caption{Pictorial view of the intuitive idea behind the refactorization process described in the main text. A codeword is a sequence of $L$ letters and, accordingly, is an element of a $2^{L\,{\cal M}}$-dimensional Hilbert space ${\cal H}_L$. Codewords are not evenly distributed in ${\cal H}_L$, but are concentrated in a typical subspace $\Lambda$ of dimension $d_\Lambda$. After enlarging ${\cal H}_L$ by a tensor product of the codewords with a pure state $\ket{0}_D$ of an ancillary $d_\Lambda$-dimensional system $D$, we have a bigger system whose states are still concentrated in a $d_\Lambda$-dimensional subspace. A unitary transformation can always be found to map this subspace into the subspace defined by the tensor product of a pure state $\ket{0}_L\in{\cal H}_L$ and ${\cal H}_D$. This space is factorizable; the first part is fed into a QIHE and the second part undergoes the decoding process to extract the information sent from the emitter.}
\label{fig::refactorizattion}
\end{figure}
Next we define the enlarged typical subspace $\Gamma:=\Lambda\, \otimes\,0^{d_\Lambda}$, where $0^{d_\Lambda}$ is the one-dimensional subspace of ${\cal H}_D$ spanned by $\ket{0}_D$.

\begin{figure*}[t]\begin{center}
\includegraphics[width=0.75\textwidth]{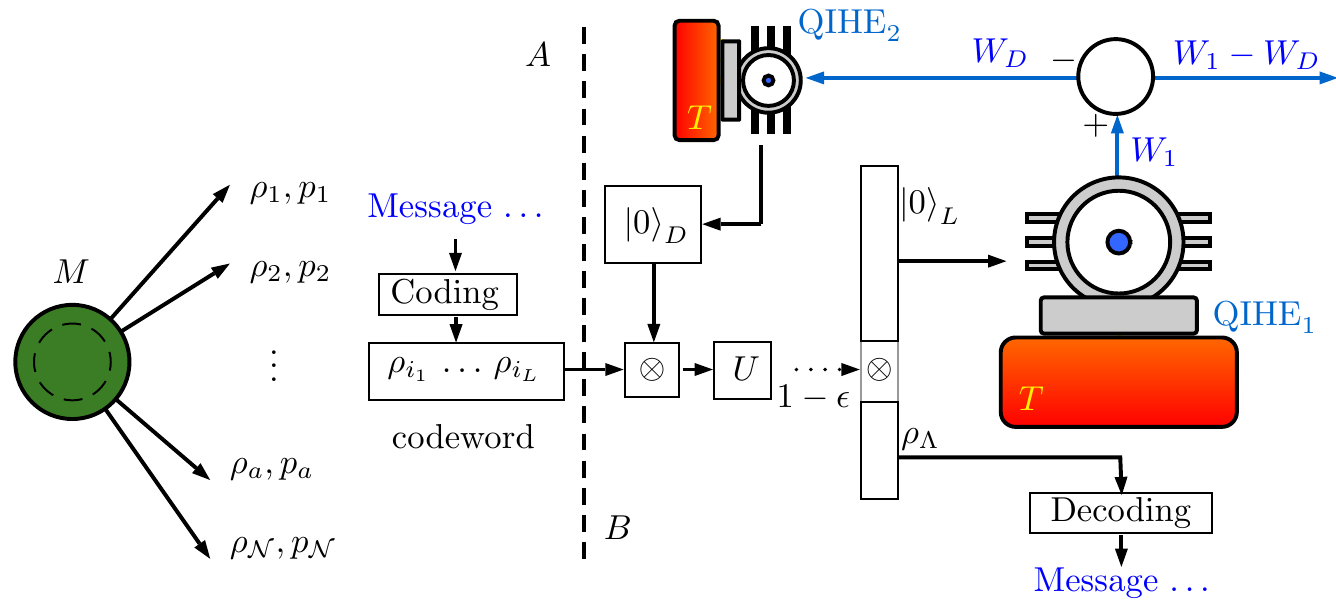}\end{center}
\caption{Schematic view of the process described in Section \ref{sec::simultaneous} for the simultaneous extraction of work and  transmission of messages. There is a source of copies of a physical system $M$ available to user $A$ (Alice). Copies are in one of a predefined set of states $\rho_a$ (letters), each with a probability $p_a$. Alice can arrange blocks of $L$ copies in predefined {\em codewords}, which have been previously agreed with user $B$ (Bob). The relative frequencies expected for each letter must follow the probability distribution. When the codeword reaches Bob, it undergoes a tensor product with a pure state $\ket{0}_D$ of an ancillary system $D$ whose dimension is equal to that of the typical subspace $\Lambda$. The enlarged codeword then undergoes a unitary transformation $U$ which, with at least $1-\epsilon$ probability, factorizes it into two parts; one of them is a pure state $\ket{0}_L$ in a $2^{L{\cal M}}$-dimensional space. The other factor is equivalent to the projection of the codeword on the typical space $\Lambda$, which can be further decoded to recover the original message. The state $\ket{0}_L$ can be fed to a QIHE (QIHE$_1$) to extract a work $W_1=k_B\,T\,(\ln 2)\,{\cal M}\,L$. Part of it, $W_D= k_B\,T\,(\ln 2)\,\log_2 d_\Lambda$, has to be used to operate another QIHE (QIHE$_2$) acting reversely to generate a pure state $\ket{0}_D$ out of completely depolarized systems.    }
\label{fig::chart}
\end{figure*}

 As the next step we define a unitary transformation $U$ which maps the enlarged typical subspace $\Gamma$ into the subspace $ \ket{0}_{L}\,\otimes\,{\cal H}_D$, where $\ket{0}_L$ is a pure state of ${\cal H}_L$. As both subspaces have the same dimension, such a unitary transformation is guaranteed to exist. This completes the process of transferring codewords in $\Lambda$ to $D$, as is depicted in Fig.\ref{fig::refactorizattion}. The first part of the system is fed to a QIHE, which will yield a work 
 \bel{wy}
 W_1=\,k_B\,T\,(\ln 2)\,L \,{\cal M}
 \eel
with a probability $>1-\epsilon$, corresponding to the case that the ${\cal H}_L$ part is in the $\ket{0}_L$ state or, equivalently, that the codeword is in $\Lambda$ and such probability is given by Eq.\ref{prob}. If not, the work may be negative, but never less than
$-W_1$. Accordingly, the average work obtained verifies
\bel{workop}
W\geq \,k_B\,T\,(\ln 2)\,L \,{\cal M}(1-2\epsilon),
\eel
from which we should substract the one necessary to restore the ancillary system
\bel{wresmu}
W_D = \,k_B\,T\,(\ln 2)\,\log_2 d_\Lambda,
\eel
and taking into account Eq.\ref{dimbound}, 
\bel{wres}
W_D \leq \,k_B\,T\,(\ln 2)\,(L S(\rho_B) + L\delta)
\eel
and consequently, the average net work per letter gained by the QIHE at $B$ verifies
\bel{ww}
W_{\rm letter} \geq \,k_B\,T\,(\ln 2)\,({\cal M}(1-2\epsilon)- S(\rho_B)(1+\delta))
\eel 
We also know, from the theory of QIHE and ultimately from the Second Principle of Thermodynamics that  
\bel{wwz}
W_{\rm letter} \leq \,k_B\,T\,(\ln 2)\,({\cal M}- S(\rho_B))
\eel 
so that the average work extractable per letter, in the $L\to\infty$ asymptotic limit, reads
\bel{ww}
{\cal E}:=W_{\rm letter} = \,k_B\,T\,(\ln 2)\,({\cal M}- S(\rho_B))
\eel 
and the Eq.\ref{ml} holds for the case of simultaneous QIHE and communication use of a source of quantum states. Fig.\ref{fig::chart} represents the process.
 
Given the set of states $\rho_a$ of a system $D$, with probabilities $p_a$, as in the previous paragraphs, the last question to be discussed is whether Alice and Bob can arrange a protocol to increase ${\cal E}$ and decrease ${\cal C}$ or viceversa. The answer is that ${\cal C}$ can not be increased over $\chi$, on account of Holevo's theorem, but ${\cal E}$ can be improved if Alice and Bob previously agree on decreasing ${\cal C}$. 

\begin{figure}[t]\begin{center}
\includegraphics[width=0.35\textwidth]{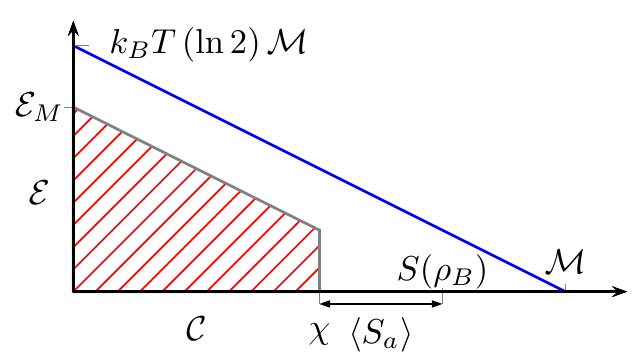}\end{center}
\caption{Graphical representation of the mutual limitation between ${\cal E}$ (extractable work per letter) and ${\cal C}$ (communication bits per letter) in the asymptotic limit. The hatched area contains the achievable values for ${\cal E},\,{\cal C}$. The maximum value of ${\cal C}$ is Holevo's bound $\chi:=S(\rho_B)-\left<\,S_a\,\right>$. If Alice and Bob agree on reducing ${\cal C}$ below this value, a higher value for ${\cal E}$ can be achieved, with a maximum of ${\cal E}_M:=k_B\,T\,(\ln 2)\,\left({\cal M}-\left<\,S_a\,\right>\right)$. The blue line represents the case when the letters are orthogonal pure states with equal probabilities. This is the situation which offers maximum possibilities to the users.}
\label{fig::recta}
\end{figure}

A way to do it would be a previous agreement on always sending blocks of $n$ copies of each letter. That is equivalent to replacing the alphabet of letters $\rho_a$ with probabilities $p_a$ by a new alphabet of letters $\rho'_a:=\rho_a^{\otimes\,n}$ with the same probabilities $p_a$. In this case, 
\bel{nop}
S(\rho'_a)=n\,S(\rho_a)\,,\,{\cal C'}\leq n\,{\cal C}\,,\,{\cal M}'=n {\cal M}\,,
\eel
and, thus, 
\bel{thus}
\frac{{\cal E}'}{n}={\cal M} - \frac{{\cal C}'}{n} - \left<\,S_a\,\right>
\eel
In the limit $n\to\infty$, $\frac{{\cal C}'}{n}\to 0$ and consequently
\bel{conseq}
\frac{{\cal E}'}{n}\to{\cal M} -  \left<\,S_a\,\right>
\eel
This value can also be reached if Alice and Bob arrange a particular ordering of the states, declining to use them for  communication. Fig.\ref{fig::recta} represents the simultaneous values per letter of 1) the extractable work ${\cal E}$ and 2) communication bits ${\cal C}$ that can be attained when Alice disposes over a supply of copies of a physical system $M$ that can be transferred to Bob. 

\section{\label{sec::results}{Results and Conclusions}}

New domains for the application of quantum information theory, especially cryptography and entanglement have been presented. It is assumed that users have access to suitable QIHEs. In particular, this paper describes protocols for work extraction from a single thermal bath through the distribution of messenger qubits, with increasingly complex features. Procedures for requiring collaboration from other users to unlock work extraction are also presented using both strongly entangled and classically correlated multipartite quantum ancillas.  Specifically, the following possibilities have been presented:
\begin{enumerate}
\item Simple transmission of messenger systems whose state is not completely depolarized for the receiver. He can extract a work equal to $k_B\,T\,(\ln 2)\,(S_{\sc max} - S(\rho))$.
\item Encrypted transmission of messenger systems though the use of previously entangled bipartite systems. This technique makes the transmitted system useless for illegitimate users that might intercept them. Quantum systems prove to be able to supply twice as much work as classical ones, because of the same physics that is behind the feature of superdense coding in quantum communication protocols.
\item In a multi-user scenario, where users initially share generalized GHZ states, any users can enable all the other ones to  extract work.
\item Also in a multi-user environment, if some correlated classical states are shared among all users, all but one can enable  the other to extract work.
 \end{enumerate}

In order to find a relation between the two possible uses of messenger systems, a mutual limitation between communication and energy has been derived in section \ref{sec::limitation} leading to Eq.\ref{ml}. This result seems quite natural, from the complementary perspectives of gaining certainty (communication) and increasing uncertainty (traded for work). In section \ref{sec::simultaneous} the validity of Eq.\ref{ml} is generalized to the case of simultaneously using the messenger system for communication and work extraction. The results are represented in Fig.\ref{fig::recta}. It shows that the communication capacity ${\cal C}$ cannot be increased over the Holevo's bound, although it can be arbitrarily reduced, upon agreement of both users, to improve the work extraction capacity up to $k_B\,T\,(\ln 2)\,\left({\cal M} - \left< S_a \right>\right)$.

\begin{acknowledgments}

We thank the Spanish MINECO grant FIS2012-33152, FIS2015-67411, the CAM research consortium QUITEMAD+ S2013/ICE-2801, the U.S. Army Research Office through grant W911NF-14-1-0103 for partial financial support

\end{acknowledgments}

\bibliography{a-biblio-cryptoenergy}

\providecommand{\noopsort}[1]{}\providecommand{\singleletter}[1]{#1}
\begin{thebibliography}{63}%
\makeatletter
\providecommand \@ifxundefined [1]{%
 \@ifx{#1\undefined}
}%
\providecommand \@ifnum [1]{%
 \ifnum #1\expandafter \@firstoftwo
 \else \expandafter \@secondoftwo
 \fi
}%
\providecommand \@ifx [1]{%
 \ifx #1\expandafter \@firstoftwo
 \else \expandafter \@secondoftwo
 \fi
}%
\providecommand \natexlab [1]{#1}%
\providecommand \enquote  [1]{``#1''}%
\providecommand \bibnamefont  [1]{#1}%
\providecommand \bibfnamefont [1]{#1}%
\providecommand \citenamefont [1]{#1}%
\providecommand \href@noop [0]{\@secondoftwo}%
\providecommand \href [0]{\begingroup \@sanitize@url \@href}%
\providecommand \@href[1]{\@@startlink{#1}\@@href}%
\providecommand \@@href[1]{\endgroup#1\@@endlink}%
\providecommand \@sanitize@url [0]{\catcode `\\12\catcode `\$12\catcode
  `\&12\catcode `\#12\catcode `\^12\catcode `\_12\catcode `\%12\relax}%
\providecommand \@@startlink[1]{}%
\providecommand \@@endlink[0]{}%
\providecommand \url  [0]{\begingroup\@sanitize@url \@url }%
\providecommand \@url [1]{\endgroup\@href {#1}{\urlprefix }}%
\providecommand \urlprefix  [0]{URL }%
\providecommand \Eprint [0]{\href }%
\providecommand \doibase [0]{http://dx.doi.org/}%
\providecommand \selectlanguage [0]{\@gobble}%
\providecommand \bibinfo  [0]{\@secondoftwo}%
\providecommand \bibfield  [0]{\@secondoftwo}%
\providecommand \translation [1]{[#1]}%
\providecommand \BibitemOpen [0]{}%
\providecommand \bibitemStop [0]{}%
\providecommand \bibitemNoStop [0]{.\EOS\space}%
\providecommand \EOS [0]{\spacefactor3000\relax}%
\providecommand \BibitemShut  [1]{\csname bibitem#1\endcsname}%
\let\auto@bib@innerbib\@empty
\bibitem [{\citenamefont {Landauer}(1961)}]{Landauer1961}%
  \BibitemOpen
  \bibfield  {author} {\bibinfo {author} {\bibfnamefont {R.}~\bibnamefont
  {Landauer}},\ }\href {\doibase 10.1147/rd.53.0183} {\bibfield  {journal}
  {\bibinfo  {journal} {IBM J. Res. Dev.}\ }\textbf {\bibinfo {volume} {5}},\
  \bibinfo {pages} {183} (\bibinfo {year} {1961})}\BibitemShut {NoStop}%
\bibitem [{\citenamefont {B\'{e}rut}\ \emph {et~al.}(2012)\citenamefont
  {B\'{e}rut}, \citenamefont {Arakelyan}, \citenamefont {Petrosyan},
  \citenamefont {Ciliberto}, \citenamefont {Dillenschneider},\ and\
  \citenamefont {Lutz}}]{Berut2012}%
  \BibitemOpen
  \bibfield  {author} {\bibinfo {author} {\bibfnamefont {A.}~\bibnamefont
  {B\'{e}rut}}, \bibinfo {author} {\bibfnamefont {A.}~\bibnamefont
  {Arakelyan}}, \bibinfo {author} {\bibfnamefont {A.}~\bibnamefont
  {Petrosyan}}, \bibinfo {author} {\bibfnamefont {S.}~\bibnamefont
  {Ciliberto}}, \bibinfo {author} {\bibfnamefont {R.}~\bibnamefont
  {Dillenschneider}}, \ and\ \bibinfo {author} {\bibfnamefont {E.}~\bibnamefont
  {Lutz}},\ }\href {\doibase 10.1038/nature10872} {\bibfield  {journal}
  {\bibinfo  {journal} {Nature}\ }\textbf {\bibinfo {volume} {483}},\ \bibinfo
  {pages} {187} (\bibinfo {year} {2012})}\BibitemShut {NoStop}%
\bibitem [{\citenamefont {Bennett}(1973)}]{Bennett1973}%
  \BibitemOpen
  \bibfield  {author} {\bibinfo {author} {\bibfnamefont {C.~H.}\ \bibnamefont
  {Bennett}},\ }\href {\doibase 10.1147/rd.176.0525} {\bibfield  {journal}
  {\bibinfo  {journal} {IBM J. Res. Dev.}\ }\textbf {\bibinfo {volume} {17}},\
  \bibinfo {pages} {525} (\bibinfo {year} {1973})}\BibitemShut {NoStop}%
\bibitem [{\citenamefont {Feynman}\ and\ \citenamefont
  {Shor}(1982)}]{Feynman1982}%
  \BibitemOpen
  \bibfield  {author} {\bibinfo {author} {\bibfnamefont {R.}~\bibnamefont
  {Feynman}}\ and\ \bibinfo {author} {\bibfnamefont {P.~W.}\ \bibnamefont
  {Shor}},\ }\href@noop {} {\bibfield  {journal} {\bibinfo  {journal} {SIAM
  Journal on Computing}\ }\textbf {\bibinfo {volume} {26}},\ \bibinfo {pages}
  {1484} (\bibinfo {year} {1982})}\BibitemShut {NoStop}%
\bibitem [{\citenamefont {Benioff}(1982)}]{Benioff1984}%
  \BibitemOpen
  \bibfield  {author} {\bibinfo {author} {\bibfnamefont {P.}~\bibnamefont
  {Benioff}},\ }\href {\doibase 10.1007/BF01342185} {\bibfield  {journal}
  {\bibinfo  {journal} {Journal of Statistical Physics}\ }\textbf {\bibinfo
  {volume} {29}},\ \bibinfo {pages} {515} (\bibinfo {year} {1982})}\BibitemShut
  {NoStop}%
\bibitem [{\citenamefont {Deutsch}(1985)}]{Deutsch1985}%
  \BibitemOpen
  \bibfield  {author} {\bibinfo {author} {\bibfnamefont {D.}~\bibnamefont
  {Deutsch}},\ }\href {\doibase 10.2307/2397601} {\bibfield  {journal}
  {\bibinfo  {journal} {Proceedings of the Royal Society of London. Series A,
  Mathematical and Physical Sciences}\ }\textbf {\bibinfo {volume} {400}},\
  \bibinfo {pages} {97} (\bibinfo {year} {1985})}\BibitemShut {NoStop}%
\bibitem [{\citenamefont {Galindo}\ and\ \citenamefont
  {Martin-Delgado}(2002)}]{Galindo2002}%
  \BibitemOpen
  \bibfield  {author} {\bibinfo {author} {\bibfnamefont {A.}~\bibnamefont
  {Galindo}}\ and\ \bibinfo {author} {\bibfnamefont {M.~A.}\ \bibnamefont
  {Martin-Delgado}},\ }\href {\doibase 10.1103/RevModPhys.74.347} {\bibfield
  {journal} {\bibinfo  {journal} {Rev. Mod. Phys.}\ }\textbf {\bibinfo {volume}
  {74}},\ \bibinfo {pages} {347} (\bibinfo {year} {2002})}\BibitemShut
  {NoStop}%
\bibitem [{\citenamefont {Nielsen}\ and\ \citenamefont
  {Chuang}(2000)}]{Nielsen2000}%
  \BibitemOpen
  \bibfield  {author} {\bibinfo {author} {\bibfnamefont {M.}~\bibnamefont
  {Nielsen}}\ and\ \bibinfo {author} {\bibfnamefont {I.}~\bibnamefont
  {Chuang}},\ }\href@noop {} {\emph {\bibinfo {title} {Quantum computation and
  quantum information}}}\ (\bibinfo  {publisher} {Cambridge University Press},\
  \bibinfo {year} {2000})\BibitemShut {NoStop}%
\bibitem [{\citenamefont {Szilard}(1964)}]{Szilard1964}%
  \BibitemOpen
  \bibfield  {author} {\bibinfo {author} {\bibfnamefont {L.}~\bibnamefont
  {Szilard}},\ }\href {\doibase 10.1002/bs.3830090402} {\bibfield  {journal}
  {\bibinfo  {journal} {Behavioral Science}\ }\textbf {\bibinfo {volume} {9}},\
  \bibinfo {pages} {301} (\bibinfo {year} {1964})}\BibitemShut {NoStop}%
\bibitem [{\citenamefont {Bennett}(1987)}]{Bennett1987}%
  \BibitemOpen
  \bibfield  {author} {\bibinfo {author} {\bibfnamefont {C.~H.}\ \bibnamefont
  {Bennett}},\ }\href@noop {} {\bibfield  {journal} {\bibinfo  {journal}
  {Scientific American}\ }\textbf {\bibinfo {volume} {259}} (\bibinfo {year}
  {1987})}\BibitemShut {NoStop}%
\bibitem [{\citenamefont {Zurek}(2003{\natexlab{a}})}]{Zurek2003-1}%
  \BibitemOpen
  \bibfield  {author} {\bibinfo {author} {\bibfnamefont {W.~H.}\ \bibnamefont
  {Zurek}},\ }\href@noop {} {\enquote {\bibinfo {title} {{Maxwell's Demon},
  {Szilard's Engine} and quantum measurements},}\ }\bibinfo {howpublished}
  {eprint arXiv:quant-ph/0301076} (\bibinfo {year}
  {2003}{\natexlab{a}})\BibitemShut {NoStop}%
\bibitem [{\citenamefont {Kim}\ \emph {et~al.}(2011)\citenamefont {Kim},
  \citenamefont {Sagawa}, \citenamefont {Liberato},\ and\ \citenamefont
  {Ueda}}]{Kim2011}%
  \BibitemOpen
  \bibfield  {author} {\bibinfo {author} {\bibfnamefont {S.}~\bibnamefont
  {Kim}}, \bibinfo {author} {\bibfnamefont {T.}~\bibnamefont {Sagawa}},
  \bibinfo {author} {\bibfnamefont {S.~D.}\ \bibnamefont {Liberato}}, \ and\
  \bibinfo {author} {\bibfnamefont {M.}~\bibnamefont {Ueda}},\ }\href@noop {}
  {\bibfield  {journal} {\bibinfo  {journal} {Phys. Rev. Lett.}\ }\textbf
  {\bibinfo {volume} {106}},\ \bibinfo {pages} {070401} (\bibinfo {year}
  {2011})}\BibitemShut {NoStop}%
\bibitem [{\citenamefont {Kish}\ and\ \citenamefont
  {Granqvist}(2012)}]{Kish2012}%
  \BibitemOpen
  \bibfield  {author} {\bibinfo {author} {\bibfnamefont {L.~B.}\ \bibnamefont
  {Kish}}\ and\ \bibinfo {author} {\bibfnamefont {C.~G.}\ \bibnamefont
  {Granqvist}},\ }\href {http://stacks.iop.org/0295-5075/98/i=6/a=68001}
  {\bibfield  {journal} {\bibinfo  {journal} {EPL (Europhysics Letters)}\
  }\textbf {\bibinfo {volume} {98}},\ \bibinfo {pages} {68001} (\bibinfo {year}
  {2012})}\BibitemShut {NoStop}%
\bibitem [{\citenamefont {Horowitz}\ and\ \citenamefont
  {Parrondo}(2011)}]{Horowitz2011}%
  \BibitemOpen
  \bibfield  {author} {\bibinfo {author} {\bibfnamefont {J.~M.}\ \bibnamefont
  {Horowitz}}\ and\ \bibinfo {author} {\bibfnamefont {J.~M.~R.}\ \bibnamefont
  {Parrondo}},\ }\href {http://stacks.iop.org/1367-2630/13/i=12/a=123019}
  {\bibfield  {journal} {\bibinfo  {journal} {New Journal of Physics}\ }\textbf
  {\bibinfo {volume} {13}},\ \bibinfo {pages} {123019} (\bibinfo {year}
  {2011})}\BibitemShut {NoStop}%
\bibitem [{\citenamefont {Diaz de~la Cruz}\ and\ \citenamefont
  {Martin-Delgado}(2014)}]{Diaz2014}%
  \BibitemOpen
  \bibfield  {author} {\bibinfo {author} {\bibfnamefont {J.~M.}\ \bibnamefont
  {Diaz de~la Cruz}}\ and\ \bibinfo {author} {\bibfnamefont {M.~A.}\
  \bibnamefont {Martin-Delgado}},\ }\href {\doibase 10.1103/PhysRevA.89.032327}
  {\bibfield  {journal} {\bibinfo  {journal} {Phys. Rev. A}\ }\textbf {\bibinfo
  {volume} {89}},\ \bibinfo {pages} {032327} (\bibinfo {year}
  {2014})}\BibitemShut {NoStop}%
\bibitem [{\citenamefont {Uzdin}\ and\ \citenamefont
  {Kosloff}(2014)}]{Uzdin2014}%
  \BibitemOpen
  \bibfield  {author} {\bibinfo {author} {\bibfnamefont {R.}~\bibnamefont
  {Uzdin}}\ and\ \bibinfo {author} {\bibfnamefont {R.}~\bibnamefont
  {Kosloff}},\ }\href {http://stacks.iop.org/1367-2630/16/i=9/a=095003}
  {\bibfield  {journal} {\bibinfo  {journal} {New Journal of Physics}\ }\textbf
  {\bibinfo {volume} {16}},\ \bibinfo {pages} {095003} (\bibinfo {year}
  {2014})}\BibitemShut {NoStop}%
\bibitem [{\citenamefont {Toyabe}\ \emph {et~al.}(2010)\citenamefont {Toyabe},
  \citenamefont {Sagawa}, \citenamefont {Ueda}, \citenamefont {Muneyuki},\ and\
  \citenamefont {Sano}}]{Toyabe2010}%
  \BibitemOpen
  \bibfield  {author} {\bibinfo {author} {\bibfnamefont {S.}~\bibnamefont
  {Toyabe}}, \bibinfo {author} {\bibfnamefont {T.}~\bibnamefont {Sagawa}},
  \bibinfo {author} {\bibfnamefont {M.}~\bibnamefont {Ueda}}, \bibinfo {author}
  {\bibfnamefont {E.}~\bibnamefont {Muneyuki}}, \ and\ \bibinfo {author}
  {\bibfnamefont {M.}~\bibnamefont {Sano}},\ }\href {\doibase
  10.1038/nphys1821} {\bibfield  {journal} {\bibinfo  {journal} {Nature Phys}\
  }\textbf {\bibinfo {volume} {6}},\ \bibinfo {pages} {998} (\bibinfo {year}
  {2010})}\BibitemShut {NoStop}%
\bibitem [{\citenamefont {Scully}(2001)}]{Scully2001}%
  \BibitemOpen
  \bibfield  {author} {\bibinfo {author} {\bibfnamefont {M.~O.}\ \bibnamefont
  {Scully}},\ }\href {\doibase 10.1103/PhysRevLett.87.220601} {\bibfield
  {journal} {\bibinfo  {journal} {Phys. Rev. Lett.}\ }\textbf {\bibinfo
  {volume} {87}},\ \bibinfo {pages} {220601} (\bibinfo {year}
  {2001})}\BibitemShut {NoStop}%
\bibitem [{\citenamefont {Scully}\ \emph {et~al.}(2005)\citenamefont {Scully},
  \citenamefont {Rostovtsev}, \citenamefont {Sariyanni},\ and\ \citenamefont
  {Zubairy}}]{Scully2005}%
  \BibitemOpen
  \bibfield  {author} {\bibinfo {author} {\bibfnamefont {M.~O.}\ \bibnamefont
  {Scully}}, \bibinfo {author} {\bibfnamefont {Y.}~\bibnamefont {Rostovtsev}},
  \bibinfo {author} {\bibfnamefont {Z.-E.}\ \bibnamefont {Sariyanni}}, \ and\
  \bibinfo {author} {\bibfnamefont {M.~S.}\ \bibnamefont {Zubairy}},\ }\href
  {\doibase http://dx.doi.org/10.1016/j.physe.2005.05.046} {\bibfield
  {journal} {\bibinfo  {journal} {Physica E: Low-dimensional Systems and
  Nanostructures}\ }\textbf {\bibinfo {volume} {29}},\ \bibinfo {pages} {29 }
  (\bibinfo {year} {2005})},\ \bibinfo {note} {frontiers of QuantumProceedings
  of the International Conference Frontiers of Quantum and Mesoscopic
  Thermodynamics}\BibitemShut {NoStop}%
\bibitem [{\citenamefont {Rostovtsev}\ \emph {et~al.}(2005)\citenamefont
  {Rostovtsev}, \citenamefont {Sariyanni}, \citenamefont {Zubairy},\ and\
  \citenamefont {Scully}}]{Rostovtsev2005}%
  \BibitemOpen
  \bibfield  {author} {\bibinfo {author} {\bibfnamefont {Y.}~\bibnamefont
  {Rostovtsev}}, \bibinfo {author} {\bibfnamefont {Z.-E.}\ \bibnamefont
  {Sariyanni}}, \bibinfo {author} {\bibfnamefont {M.~S.}\ \bibnamefont
  {Zubairy}}, \ and\ \bibinfo {author} {\bibfnamefont {M.~O.}\ \bibnamefont
  {Scully}},\ }\href {\doibase http://dx.doi.org/10.1016/j.physe.2005.05.052}
  {\bibfield  {journal} {\bibinfo  {journal} {Physica E: Low-dimensional
  Systems and Nanostructures}\ }\textbf {\bibinfo {volume} {29}},\ \bibinfo
  {pages} {40 } (\bibinfo {year} {2005})},\ \bibinfo {note} {frontiers of
  QuantumProceedings of the International Conference Frontiers of Quantum and
  Mesoscopic Thermodynamics}\BibitemShut {NoStop}%
\bibitem [{\citenamefont {Sariyanni}\ \emph {et~al.}(2005)\citenamefont
  {Sariyanni}, \citenamefont {Rostovtsev}, \citenamefont {Zubairy},\ and\
  \citenamefont {Scully}}]{Sariyanni2005}%
  \BibitemOpen
  \bibfield  {author} {\bibinfo {author} {\bibfnamefont {Z.-E.}\ \bibnamefont
  {Sariyanni}}, \bibinfo {author} {\bibfnamefont {Y.}~\bibnamefont
  {Rostovtsev}}, \bibinfo {author} {\bibfnamefont {M.~S.}\ \bibnamefont
  {Zubairy}}, \ and\ \bibinfo {author} {\bibfnamefont {M.~O.}\ \bibnamefont
  {Scully}},\ }\href {\doibase http://dx.doi.org/10.1016/j.physe.2005.05.045}
  {\bibfield  {journal} {\bibinfo  {journal} {Physica E: Low-dimensional
  Systems and Nanostructures}\ }\textbf {\bibinfo {volume} {29}},\ \bibinfo
  {pages} {47 } (\bibinfo {year} {2005})},\ \bibinfo {note} {frontiers of
  QuantumProceedings of the International Conference Frontiers of Quantum and
  Mesoscopic Thermodynamics}\BibitemShut {NoStop}%
\bibitem [{\citenamefont {Quan}\ \emph {et~al.}(2006)\citenamefont {Quan},
  \citenamefont {Wang}, \citenamefont {Liu}, \citenamefont {Sun},\ and\
  \citenamefont {Nori}}]{Quan2006}%
  \BibitemOpen
  \bibfield  {author} {\bibinfo {author} {\bibfnamefont {H.~T.}\ \bibnamefont
  {Quan}}, \bibinfo {author} {\bibfnamefont {Y.~D.}\ \bibnamefont {Wang}},
  \bibinfo {author} {\bibfnamefont {Y.-x.}\ \bibnamefont {Liu}}, \bibinfo
  {author} {\bibfnamefont {C.~P.}\ \bibnamefont {Sun}}, \ and\ \bibinfo
  {author} {\bibfnamefont {F.}~\bibnamefont {Nori}},\ }\href {\doibase
  10.1103/PhysRevLett.97.180402} {\bibfield  {journal} {\bibinfo  {journal}
  {Phys. Rev. Lett.}\ }\textbf {\bibinfo {volume} {97}},\ \bibinfo {pages}
  {180402} (\bibinfo {year} {2006})}\BibitemShut {NoStop}%
\bibitem [{\citenamefont {del Rio}\ \emph {et~al.}(2011)\citenamefont {del
  Rio}, \citenamefont {Aberg}, \citenamefont {Renner}, \citenamefont
  {Dahlsten},\ and\ \citenamefont {Vedral}}]{Delrio2011}%
  \BibitemOpen
  \bibfield  {author} {\bibinfo {author} {\bibfnamefont {L.}~\bibnamefont {del
  Rio}}, \bibinfo {author} {\bibfnamefont {J.}~\bibnamefont {Aberg}}, \bibinfo
  {author} {\bibfnamefont {R.}~\bibnamefont {Renner}}, \bibinfo {author}
  {\bibfnamefont {O.}~\bibnamefont {Dahlsten}}, \ and\ \bibinfo {author}
  {\bibfnamefont {V.}~\bibnamefont {Vedral}},\ }\href@noop {} {\enquote
  {\bibinfo {title} {The thermodynamic meaning of negative entropy},}\ }
  (\bibinfo {year} {2011}),\ \Eprint {http://arxiv.org/abs/1009.1630v2}
  {arXiv:1009.1630v2} \BibitemShut {NoStop}%
\bibitem [{\citenamefont {Plesch}\ \emph
  {et~al.}(2013{\natexlab{a}})\citenamefont {Plesch}, \citenamefont {Dahlsten},
  \citenamefont {Goold},\ and\ \citenamefont {Vedral}}]{Plesch2013}%
  \BibitemOpen
  \bibfield  {author} {\bibinfo {author} {\bibfnamefont {M.}~\bibnamefont
  {Plesch}}, \bibinfo {author} {\bibfnamefont {O.}~\bibnamefont {Dahlsten}},
  \bibinfo {author} {\bibfnamefont {J.}~\bibnamefont {Goold}}, \ and\ \bibinfo
  {author} {\bibfnamefont {V.}~\bibnamefont {Vedral}},\ }\href {\doibase
  10.1103/PhysRevLett.111.188901} {\bibfield  {journal} {\bibinfo  {journal}
  {Phys. Rev. Lett.}\ }\textbf {\bibinfo {volume} {111}},\ \bibinfo {pages}
  {188901} (\bibinfo {year} {2013}{\natexlab{a}})}\BibitemShut {NoStop}%
\bibitem [{\citenamefont {Funo}\ \emph {et~al.}(2013)\citenamefont {Funo},
  \citenamefont {Watanabe},\ and\ \citenamefont {Ueda}}]{Funo2013}%
  \BibitemOpen
  \bibfield  {author} {\bibinfo {author} {\bibfnamefont {K.}~\bibnamefont
  {Funo}}, \bibinfo {author} {\bibfnamefont {Y.}~\bibnamefont {Watanabe}}, \
  and\ \bibinfo {author} {\bibfnamefont {M.}~\bibnamefont {Ueda}},\ }\href
  {\doibase 10.1103/PhysRevA.88.052319} {\bibfield  {journal} {\bibinfo
  {journal} {Phys. Rev. A}\ }\textbf {\bibinfo {volume} {88}},\ \bibinfo
  {pages} {052319} (\bibinfo {year} {2013})}\BibitemShut {NoStop}%
\bibitem [{\citenamefont {Piechocinska}(2000)}]{Piechocinska2000}%
  \BibitemOpen
  \bibfield  {author} {\bibinfo {author} {\bibfnamefont {B.}~\bibnamefont
  {Piechocinska}},\ }\href {\doibase 10.1103/PhysRevA.61.062314} {\bibfield
  {journal} {\bibinfo  {journal} {Phys. Rev. A}\ }\textbf {\bibinfo {volume}
  {61}},\ \bibinfo {pages} {062314} (\bibinfo {year} {2000})}\BibitemShut
  {NoStop}%
\bibitem [{\citenamefont {Zhou}\ and\ \citenamefont {Segal}(2010)}]{Zhou2010}%
  \BibitemOpen
  \bibfield  {author} {\bibinfo {author} {\bibfnamefont {Y.}~\bibnamefont
  {Zhou}}\ and\ \bibinfo {author} {\bibfnamefont {D.}~\bibnamefont {Segal}},\
  }\href {\doibase 10.1103/PhysRevE.82.011120} {\bibfield  {journal} {\bibinfo
  {journal} {Phys. Rev. E}\ }\textbf {\bibinfo {volume} {82}},\ \bibinfo
  {pages} {011120} (\bibinfo {year} {2010})}\BibitemShut {NoStop}%
\bibitem [{\citenamefont {Lloyd}(1997)}]{Lloyd1997}%
  \BibitemOpen
  \bibfield  {author} {\bibinfo {author} {\bibfnamefont {S.}~\bibnamefont
  {Lloyd}},\ }\href {\doibase 10.1103/PhysRevA.56.3374} {\bibfield  {journal}
  {\bibinfo  {journal} {Phys. Rev. A}\ }\textbf {\bibinfo {volume} {56}},\
  \bibinfo {pages} {3374} (\bibinfo {year} {1997})}\BibitemShut {NoStop}%
\bibitem [{\citenamefont {Vidrighin}\ \emph {et~al.}(2016)\citenamefont
  {Vidrighin}, \citenamefont {Dahlsten}, \citenamefont {Barbieri},
  \citenamefont {Kim}, \citenamefont {Vedral},\ and\ \citenamefont
  {Walmsley}}]{Vedral2016}%
  \BibitemOpen
  \bibfield  {author} {\bibinfo {author} {\bibfnamefont {M.~D.}\ \bibnamefont
  {Vidrighin}}, \bibinfo {author} {\bibfnamefont {O.}~\bibnamefont {Dahlsten}},
  \bibinfo {author} {\bibfnamefont {M.}~\bibnamefont {Barbieri}}, \bibinfo
  {author} {\bibfnamefont {M.~S.}\ \bibnamefont {Kim}}, \bibinfo {author}
  {\bibfnamefont {V.}~\bibnamefont {Vedral}}, \ and\ \bibinfo {author}
  {\bibfnamefont {I.~A.}\ \bibnamefont {Walmsley}},\ }\href {\doibase
  10.1103/PhysRevLett.116.050401} {\bibfield  {journal} {\bibinfo  {journal}
  {Phys. Rev. Lett.}\ }\textbf {\bibinfo {volume} {116}},\ \bibinfo {pages}
  {050401} (\bibinfo {year} {2016})}\BibitemShut {NoStop}%
\bibitem [{\citenamefont {Oppenheim}\ \emph {et~al.}(2002)\citenamefont
  {Oppenheim}, \citenamefont {Horodecki}, \citenamefont {Horodecki},\ and\
  \citenamefont {Horodecki}}]{Oppenheim2002}%
  \BibitemOpen
  \bibfield  {author} {\bibinfo {author} {\bibfnamefont {J.}~\bibnamefont
  {Oppenheim}}, \bibinfo {author} {\bibfnamefont {M.}~\bibnamefont
  {Horodecki}}, \bibinfo {author} {\bibfnamefont {P.}~\bibnamefont
  {Horodecki}}, \ and\ \bibinfo {author} {\bibfnamefont {R.}~\bibnamefont
  {Horodecki}},\ }\href {\doibase 10.1103/PhysRevLett.89.180402} {\bibfield
  {journal} {\bibinfo  {journal} {Phys. Rev. Lett.}\ }\textbf {\bibinfo
  {volume} {89}},\ \bibinfo {pages} {180402} (\bibinfo {year}
  {2002})}\BibitemShut {NoStop}%
\bibitem [{\citenamefont {Park}\ \emph {et~al.}(2013)\citenamefont {Park},
  \citenamefont {Kim}, \citenamefont {Sagawa},\ and\ \citenamefont
  {Kim}}]{Park2013}%
  \BibitemOpen
  \bibfield  {author} {\bibinfo {author} {\bibfnamefont {J.}~\bibnamefont
  {Park}}, \bibinfo {author} {\bibfnamefont {K.}~\bibnamefont {Kim}}, \bibinfo
  {author} {\bibfnamefont {T.}~\bibnamefont {Sagawa}}, \ and\ \bibinfo {author}
  {\bibfnamefont {S.}~\bibnamefont {Kim}},\ }\href@noop {} {\enquote {\bibinfo
  {title} {Heat engine driven by purely quantum information},}\ } (\bibinfo
  {year} {2013}),\ \Eprint {http://arxiv.org/abs/1302.3011} {arXiv:1302.3011}
  \BibitemShut {NoStop}%
\bibitem [{\citenamefont {Ollivier}\ and\ \citenamefont
  {Zurek}(2002)}]{Ollivier2002}%
  \BibitemOpen
  \bibfield  {author} {\bibinfo {author} {\bibfnamefont {H.}~\bibnamefont
  {Ollivier}}\ and\ \bibinfo {author} {\bibfnamefont {W.}~\bibnamefont
  {Zurek}},\ }\href@noop {} {\bibfield  {journal} {\bibinfo  {journal} {Phys.
  Rev. Lett.}\ }\textbf {\bibinfo {volume} {88}},\ \bibinfo {pages} {017901}
  (\bibinfo {year} {2002})}\BibitemShut {NoStop}%
\bibitem [{\citenamefont {Zurek}(2003{\natexlab{b}})}]{Zurek2003-2}%
  \BibitemOpen
  \bibfield  {author} {\bibinfo {author} {\bibfnamefont {W.~H.}\ \bibnamefont
  {Zurek}},\ }\href {\doibase 10.1103/PhysRevA.67.012320} {\bibfield  {journal}
  {\bibinfo  {journal} {Phys. Rev. A}\ }\textbf {\bibinfo {volume} {67}},\
  \bibinfo {pages} {012320} (\bibinfo {year} {2003}{\natexlab{b}})}\BibitemShut
  {NoStop}%
\bibitem [{\citenamefont {Hasegawa}\ \emph {et~al.}(2010)\citenamefont
  {Hasegawa}, \citenamefont {Ishikawa}, \citenamefont {Takara},\ and\
  \citenamefont {Driebe}}]{Hasegawa2010}%
  \BibitemOpen
  \bibfield  {author} {\bibinfo {author} {\bibfnamefont {H.}~\bibnamefont
  {Hasegawa}}, \bibinfo {author} {\bibfnamefont {J.}~\bibnamefont {Ishikawa}},
  \bibinfo {author} {\bibfnamefont {K.}~\bibnamefont {Takara}}, \ and\ \bibinfo
  {author} {\bibfnamefont {D.}~\bibnamefont {Driebe}},\ }\href@noop {}
  {\bibfield  {journal} {\bibinfo  {journal} {Phys. Lett. A}\ }\textbf
  {\bibinfo {volume} {374,8}},\ \bibinfo {pages} {1001} (\bibinfo {year}
  {2010})}\BibitemShut {NoStop}%
\bibitem [{\citenamefont {Vedral}(2002)}]{Vedral2002}%
  \BibitemOpen
  \bibfield  {author} {\bibinfo {author} {\bibfnamefont {V.}~\bibnamefont
  {Vedral}},\ }\href {\doibase 10.1103/RevModPhys.74.197} {\bibfield  {journal}
  {\bibinfo  {journal} {Rev. Mod. Phys.}\ }\textbf {\bibinfo {volume} {74}},\
  \bibinfo {pages} {197} (\bibinfo {year} {2002})}\BibitemShut {NoStop}%
\bibitem [{\citenamefont {Sagawa}(2012)}]{Sagawa2012}%
  \BibitemOpen
  \bibfield  {author} {\bibinfo {author} {\bibfnamefont {T.}~\bibnamefont
  {Sagawa}},\ }\href@noop {} {\enquote {\bibinfo {title} {Second law-like
  inequalities with quantum relative entropy: An introduction},}\ } (\bibinfo
  {year} {2012}),\ \Eprint {http://arxiv.org/abs/1202.0983} {arXiv:1202.0983}
  \BibitemShut {NoStop}%
\bibitem [{\citenamefont {Brandao}\ \emph {et~al.}(2015)\citenamefont
  {Brandao}, \citenamefont {Horodecki}, \citenamefont {Ng}, \citenamefont
  {Oppenheim},\ and\ \citenamefont {Wehner}}]{Brandao2015}%
  \BibitemOpen
  \bibfield  {author} {\bibinfo {author} {\bibfnamefont {F.}~\bibnamefont
  {Brandao}}, \bibinfo {author} {\bibfnamefont {M.}~\bibnamefont {Horodecki}},
  \bibinfo {author} {\bibfnamefont {N.}~\bibnamefont {Ng}}, \bibinfo {author}
  {\bibfnamefont {J.}~\bibnamefont {Oppenheim}}, \ and\ \bibinfo {author}
  {\bibfnamefont {S.}~\bibnamefont {Wehner}},\ }\href {\doibase
  10.1073/pnas.1411728112} {\bibfield  {journal} {\bibinfo  {journal}
  {Proceedings of the National Academy of Sciences}\ }\textbf {\bibinfo
  {volume} {112}},\ \bibinfo {pages} {3275} (\bibinfo {year} {2015})},\ \Eprint
  {http://arxiv.org/abs/http://www.pnas.org/content/112/11/3275.full.pdf}
  {http://www.pnas.org/content/112/11/3275.full.pdf} \BibitemShut {NoStop}%
\bibitem [{\citenamefont {Zheng}\ and\ \citenamefont
  {Poletti}(2015)}]{YuanjianZheng2015}%
  \BibitemOpen
  \bibfield  {author} {\bibinfo {author} {\bibfnamefont {Y.}~\bibnamefont
  {Zheng}}\ and\ \bibinfo {author} {\bibfnamefont {D.}~\bibnamefont
  {Poletti}},\ }\href {\doibase 10.1103/PhysRevE.92.012110} {\bibfield
  {journal} {\bibinfo  {journal} {Phys. Rev. E}\ }\textbf {\bibinfo {volume}
  {92}},\ \bibinfo {pages} {012110} (\bibinfo {year} {2015})}\BibitemShut
  {NoStop}%
\bibitem [{\citenamefont {Plesch}\ \emph
  {et~al.}(2013{\natexlab{b}})\citenamefont {Plesch}, \citenamefont {Dahlsten},
  \citenamefont {Goold},\ and\ \citenamefont {Vedral}}]{Plesch2011}%
  \BibitemOpen
  \bibfield  {author} {\bibinfo {author} {\bibfnamefont {M.}~\bibnamefont
  {Plesch}}, \bibinfo {author} {\bibfnamefont {O.}~\bibnamefont {Dahlsten}},
  \bibinfo {author} {\bibfnamefont {J.}~\bibnamefont {Goold}}, \ and\ \bibinfo
  {author} {\bibfnamefont {V.}~\bibnamefont {Vedral}},\ }\href {\doibase
  10.1103/PhysRevLett.111.188901} {\bibfield  {journal} {\bibinfo  {journal}
  {Phys. Rev. Lett.}\ }\textbf {\bibinfo {volume} {111}},\ \bibinfo {pages}
  {188901} (\bibinfo {year} {2013}{\natexlab{b}})}\BibitemShut {NoStop}%
\bibitem [{\citenamefont {Jaramillo}\ \emph {et~al.}(2016)\citenamefont
  {Jaramillo}, \citenamefont {Beau},\ and\ \citenamefont {del
  Campo}}]{Jaramillo2015}%
  \BibitemOpen
  \bibfield  {author} {\bibinfo {author} {\bibfnamefont {J.}~\bibnamefont
  {Jaramillo}}, \bibinfo {author} {\bibfnamefont {M.}~\bibnamefont {Beau}}, \
  and\ \bibinfo {author} {\bibfnamefont {A.}~\bibnamefont {del Campo}},\ }\href
  {http://stacks.iop.org/1367-2630/18/i=7/a=075019} {\bibfield  {journal}
  {\bibinfo  {journal} {New Journal of Physics}\ }\textbf {\bibinfo {volume}
  {18}},\ \bibinfo {pages} {075019} (\bibinfo {year} {2016})}\BibitemShut
  {NoStop}%
\bibitem [{\citenamefont {Beau}\ \emph {et~al.}(2016)\citenamefont {Beau},
  \citenamefont {Jaramillo},\ and\ \citenamefont {del Campo}}]{Beau2016}%
  \BibitemOpen
  \bibfield  {author} {\bibinfo {author} {\bibfnamefont {M.}~\bibnamefont
  {Beau}}, \bibinfo {author} {\bibfnamefont {J.}~\bibnamefont {Jaramillo}}, \
  and\ \bibinfo {author} {\bibfnamefont {A.}~\bibnamefont {del Campo}},\ }\href
  {\doibase 10.3390/e18050168} {\bibfield  {journal} {\bibinfo  {journal}
  {Entropy}\ }\textbf {\bibinfo {volume} {18}},\ \bibinfo {pages} {168}
  (\bibinfo {year} {2016})}\BibitemShut {NoStop}%
\bibitem [{\citenamefont {Jeon}\ and\ \citenamefont {Kim}(2016)}]{Jeon2016}%
  \BibitemOpen
  \bibfield  {author} {\bibinfo {author} {\bibfnamefont {H.~J.}\ \bibnamefont
  {Jeon}}\ and\ \bibinfo {author} {\bibfnamefont {S.~W.}\ \bibnamefont {Kim}},\
  }\href {\doibase 10.1088/1367-2630/18/4/043002} {\bibfield  {journal}
  {\bibinfo  {journal} {New Journal of Physics}\ }\textbf {\bibinfo {volume}
  {18}},\ \bibinfo {pages} {043002} (\bibinfo {year} {2016})}\BibitemShut
  {NoStop}%
\bibitem [{\citenamefont {Kosloff}\ and\ \citenamefont
  {Feldmann}(2010)}]{Kosloff2010}%
  \BibitemOpen
  \bibfield  {author} {\bibinfo {author} {\bibfnamefont {R.}~\bibnamefont
  {Kosloff}}\ and\ \bibinfo {author} {\bibfnamefont {T.}~\bibnamefont
  {Feldmann}},\ }\href {\doibase 10.1103/PhysRevE.82.011134} {\bibfield
  {journal} {\bibinfo  {journal} {Phys. Rev. E}\ }\textbf {\bibinfo {volume}
  {82}},\ \bibinfo {pages} {011134} (\bibinfo {year} {2010})}\BibitemShut
  {NoStop}%
\bibitem [{\citenamefont {Beretta}(2012)}]{Beretta2012}%
  \BibitemOpen
  \bibfield  {author} {\bibinfo {author} {\bibfnamefont {G.~P.}\ \bibnamefont
  {Beretta}},\ }\href {http://stacks.iop.org/0295-5075/99/i=2/a=20005}
  {\bibfield  {journal} {\bibinfo  {journal} {EPL (Europhysics Letters)}\
  }\textbf {\bibinfo {volume} {99}},\ \bibinfo {pages} {20005} (\bibinfo {year}
  {2012})}\BibitemShut {NoStop}%
\bibitem [{\citenamefont {Gordon}\ \emph {et~al.}(2009)\citenamefont {Gordon},
  \citenamefont {Bensky}, \citenamefont {Gelbwaser-Klimovsky}, \citenamefont
  {Rao}, \citenamefont {Erez},\ and\ \citenamefont {Kurizki}}]{Gordon2009}%
  \BibitemOpen
  \bibfield  {author} {\bibinfo {author} {\bibfnamefont {G.}~\bibnamefont
  {Gordon}}, \bibinfo {author} {\bibfnamefont {G.}~\bibnamefont {Bensky}},
  \bibinfo {author} {\bibfnamefont {D.}~\bibnamefont {Gelbwaser-Klimovsky}},
  \bibinfo {author} {\bibfnamefont {D.~D.~B.}\ \bibnamefont {Rao}}, \bibinfo
  {author} {\bibfnamefont {N.}~\bibnamefont {Erez}}, \ and\ \bibinfo {author}
  {\bibfnamefont {G.}~\bibnamefont {Kurizki}},\ }\href
  {http://stacks.iop.org/1367-2630/11/i=12/a=123025} {\bibfield  {journal}
  {\bibinfo  {journal} {New Journal of Physics}\ }\textbf {\bibinfo {volume}
  {11}},\ \bibinfo {pages} {123025} (\bibinfo {year} {2009})}\BibitemShut
  {NoStop}%
\bibitem [{\citenamefont {Dahlsten}\ \emph {et~al.}(2011)\citenamefont
  {Dahlsten}, \citenamefont {Renner}, \citenamefont {Rieper},\ and\
  \citenamefont {Vedral}}]{Dahlsten2011}%
  \BibitemOpen
  \bibfield  {author} {\bibinfo {author} {\bibfnamefont {O.~C.~O.}\
  \bibnamefont {Dahlsten}}, \bibinfo {author} {\bibfnamefont {R.}~\bibnamefont
  {Renner}}, \bibinfo {author} {\bibfnamefont {E.}~\bibnamefont {Rieper}}, \
  and\ \bibinfo {author} {\bibfnamefont {V.}~\bibnamefont {Vedral}},\ }\href
  {http://stacks.iop.org/1367-2630/13/i=5/a=053015} {\bibfield  {journal}
  {\bibinfo  {journal} {New Journal of Physics}\ }\textbf {\bibinfo {volume}
  {13}},\ \bibinfo {pages} {053015} (\bibinfo {year} {2011})}\BibitemShut
  {NoStop}%
\bibitem [{\citenamefont {Horodecki}\ and\ \citenamefont
  {Oppenheim}(2011)}]{Horodecki2011}%
  \BibitemOpen
  \bibfield  {author} {\bibinfo {author} {\bibfnamefont {M.}~\bibnamefont
  {Horodecki}}\ and\ \bibinfo {author} {\bibfnamefont {J.}~\bibnamefont
  {Oppenheim}},\ }\href {http://arxiv.org/abs/1111.3834} {\enquote {\bibinfo
  {title} {Fundamental limitations for quantum and nano thermodynamics},}\ }
  (\bibinfo {year} {2011}),\ \Eprint {http://arxiv.org/abs/1111.3834}
  {arXiv:1111.3834} \BibitemShut {NoStop}%
\bibitem [{\citenamefont {Egloff}\ \emph {et~al.}(2013)\citenamefont {Egloff},
  \citenamefont {Dahlsten}, \citenamefont {Renner},\ and\ \citenamefont
  {Vedral}}]{Egloff2013}%
  \BibitemOpen
  \bibfield  {author} {\bibinfo {author} {\bibfnamefont {D.}~\bibnamefont
  {Egloff}}, \bibinfo {author} {\bibfnamefont {O.~C.}\ \bibnamefont
  {Dahlsten}}, \bibinfo {author} {\bibfnamefont {R.}~\bibnamefont {Renner}}, \
  and\ \bibinfo {author} {\bibfnamefont {V.}~\bibnamefont {Vedral}},\ }\href
  {http://arxiv.org/abs/1207.0434} {\  (\bibinfo {year} {2013})},\ \Eprint
  {http://arxiv.org/abs/1207.0434v2} {arXiv:1207.0434v2} \BibitemShut {NoStop}%
\bibitem [{\citenamefont {Hotta}(2008)}]{Hotta2008pla}%
  \BibitemOpen
  \bibfield  {author} {\bibinfo {author} {\bibfnamefont {M.}~\bibnamefont
  {Hotta}},\ }\href {\doibase http://dx.doi.org/10.1016/j.physleta.2008.07.007}
  {\bibfield  {journal} {\bibinfo  {journal} {Physics Letters A}\ }\textbf
  {\bibinfo {volume} {372}},\ \bibinfo {pages} {5671 } (\bibinfo {year}
  {2008})}\BibitemShut {NoStop}%
\bibitem [{\citenamefont {Hotta}(2009)}]{Hotta2009}%
  \BibitemOpen
  \bibfield  {author} {\bibinfo {author} {\bibfnamefont {M.}~\bibnamefont
  {Hotta}},\ }\href {\doibase 10.1103/PhysRevA.80.042323} {\bibfield  {journal}
  {\bibinfo  {journal} {Phys. Rev. A}\ }\textbf {\bibinfo {volume} {80}},\
  \bibinfo {pages} {042323} (\bibinfo {year} {2009})}\BibitemShut {NoStop}%
\bibitem [{\citenamefont {Trevison}\ and\ \citenamefont
  {Hotta}(2015)}]{Trevison2015}%
  \BibitemOpen
  \bibfield  {author} {\bibinfo {author} {\bibfnamefont {J.}~\bibnamefont
  {Trevison}}\ and\ \bibinfo {author} {\bibfnamefont {M.}~\bibnamefont
  {Hotta}},\ }\href {http://stacks.iop.org/1751-8121/48/i=17/a=175302}
  {\bibfield  {journal} {\bibinfo  {journal} {Journal of Physics A:
  Mathematical and Theoretical}\ }\textbf {\bibinfo {volume} {48}},\ \bibinfo
  {pages} {175302} (\bibinfo {year} {2015})}\BibitemShut {NoStop}%
\bibitem [{Note1()}]{Note1}%
  \BibitemOpen
  \bibinfo {note} {In this work, a completely depolarized system is equivalent
  to a system with maximum entropy. Taking into account that Information Heat
  Engines extract work by increasing the entropy of the fueling system, a
  completely depolarized state yields zero work.}\BibitemShut {Stop}%
\bibitem [{\citenamefont {Ma}\ \emph {et~al.}(2012)\citenamefont {Ma},
  \citenamefont {Herbst}, \citenamefont {Scheidl}, \citenamefont {Wang},
  \citenamefont {Kropatschek}, \citenamefont {rand B.~Wittmann}, \citenamefont
  {Mech}, \citenamefont {Kofler}, \citenamefont {Anisimova}, \citenamefont
  {Makarov}, \citenamefont {Jennewein}, \citenamefont {Ursin},\ and\
  \citenamefont {Zeilinger}}]{Xiao2012}%
  \BibitemOpen
  \bibfield  {author} {\bibinfo {author} {\bibfnamefont {X.}~\bibnamefont
  {Ma}}, \bibinfo {author} {\bibfnamefont {T.}~\bibnamefont {Herbst}}, \bibinfo
  {author} {\bibfnamefont {T.}~\bibnamefont {Scheidl}}, \bibinfo {author}
  {\bibfnamefont {D.}~\bibnamefont {Wang}}, \bibinfo {author} {\bibfnamefont
  {S.}~\bibnamefont {Kropatschek}}, \bibinfo {author} {\bibfnamefont {W.~N.}\
  \bibnamefont {rand B.~Wittmann}}, \bibinfo {author} {\bibfnamefont
  {A.}~\bibnamefont {Mech}}, \bibinfo {author} {\bibfnamefont {J.}~\bibnamefont
  {Kofler}}, \bibinfo {author} {\bibfnamefont {E.}~\bibnamefont {Anisimova}},
  \bibinfo {author} {\bibfnamefont {V.}~\bibnamefont {Makarov}}, \bibinfo
  {author} {\bibfnamefont {T.}~\bibnamefont {Jennewein}}, \bibinfo {author}
  {\bibfnamefont {R.}~\bibnamefont {Ursin}}, \ and\ \bibinfo {author}
  {\bibfnamefont {A.}~\bibnamefont {Zeilinger}},\ }\href@noop {} {\ \textbf
  {\bibinfo {volume} {489}},\ \bibinfo {pages} {269} (\bibinfo {year}
  {2012})}\BibitemShut {NoStop}%
\bibitem [{\citenamefont {Herbst}\ \emph {et~al.}(2015)\citenamefont {Herbst},
  \citenamefont {Scheidl}, \citenamefont {Fink}, \citenamefont {Handsteiner},
  \citenamefont {Wittmann}, \citenamefont {Ursin},\ and\ \citenamefont
  {Zeilinger}}]{Herbst2015}%
  \BibitemOpen
  \bibfield  {author} {\bibinfo {author} {\bibfnamefont {T.}~\bibnamefont
  {Herbst}}, \bibinfo {author} {\bibfnamefont {T.}~\bibnamefont {Scheidl}},
  \bibinfo {author} {\bibfnamefont {M.}~\bibnamefont {Fink}}, \bibinfo {author}
  {\bibfnamefont {J.}~\bibnamefont {Handsteiner}}, \bibinfo {author}
  {\bibfnamefont {B.}~\bibnamefont {Wittmann}}, \bibinfo {author}
  {\bibfnamefont {R.}~\bibnamefont {Ursin}}, \ and\ \bibinfo {author}
  {\bibfnamefont {A.}~\bibnamefont {Zeilinger}},\ }\href {\doibase
  10.1073/pnas.1517007112} {\bibfield  {journal} {\bibinfo  {journal}
  {Proceedings of the National Academy of Sciences}\ }\textbf {\bibinfo
  {volume} {112}},\ \bibinfo {pages} {14202} (\bibinfo {year} {2015})},\
  \Eprint
  {http://arxiv.org/abs/http://www.pnas.org/content/112/46/14202.full.pdf}
  {http://www.pnas.org/content/112/46/14202.full.pdf} \BibitemShut {NoStop}%
\bibitem [{\citenamefont {Brand\~ao}\ \emph {et~al.}(2013)\citenamefont
  {Brand\~ao}, \citenamefont {Horodecki}, \citenamefont {Oppenheim},
  \citenamefont {Renes},\ and\ \citenamefont {Spekkens}}]{Brandao2013}%
  \BibitemOpen
  \bibfield  {author} {\bibinfo {author} {\bibfnamefont {F.~G. S.~L.}\
  \bibnamefont {Brand\~ao}}, \bibinfo {author} {\bibfnamefont {M.}~\bibnamefont
  {Horodecki}}, \bibinfo {author} {\bibfnamefont {J.}~\bibnamefont
  {Oppenheim}}, \bibinfo {author} {\bibfnamefont {J.~M.}\ \bibnamefont
  {Renes}}, \ and\ \bibinfo {author} {\bibfnamefont {R.~W.}\ \bibnamefont
  {Spekkens}},\ }\href {\doibase 10.1103/PhysRevLett.111.250404} {\bibfield
  {journal} {\bibinfo  {journal} {Phys. Rev. Lett.}\ }\textbf {\bibinfo
  {volume} {111}},\ \bibinfo {pages} {250404} (\bibinfo {year}
  {2013})}\BibitemShut {NoStop}%
\bibitem [{\citenamefont {Horodecki}\ and\ \citenamefont
  {Oppenheim}(2013)}]{HorodeckiOppenheim2013}%
  \BibitemOpen
  \bibfield  {author} {\bibinfo {author} {\bibfnamefont {M.}~\bibnamefont
  {Horodecki}}\ and\ \bibinfo {author} {\bibfnamefont {J.}~\bibnamefont
  {Oppenheim}},\ }\href {\doibase 10.1038/ncomms3059} {\bibfield  {journal}
  {\bibinfo  {journal} {Nat. Comm.}\ }\textbf {\bibinfo {volume} {4}} (\bibinfo
  {year} {2013}),\ 10.1038/ncomms3059}\BibitemShut {NoStop}%
\bibitem [{\citenamefont {Sagawa}\ and\ \citenamefont
  {Ueda}(2009)}]{Sagawa2009}%
  \BibitemOpen
  \bibfield  {author} {\bibinfo {author} {\bibfnamefont {T.}~\bibnamefont
  {Sagawa}}\ and\ \bibinfo {author} {\bibfnamefont {M.}~\bibnamefont {Ueda}},\
  }\href {\doibase 10.1103/PhysRevLett.102.250602} {\bibfield  {journal}
  {\bibinfo  {journal} {Phys. Rev. Lett.}\ }\textbf {\bibinfo {volume} {102}},\
  \bibinfo {pages} {250602} (\bibinfo {year} {2009})}\BibitemShut {NoStop}%
\bibitem [{\citenamefont {De~Liberato}\ and\ \citenamefont
  {Ueda}(2011)}]{DeLiberato2011}%
  \BibitemOpen
  \bibfield  {author} {\bibinfo {author} {\bibfnamefont {S.}~\bibnamefont
  {De~Liberato}}\ and\ \bibinfo {author} {\bibfnamefont {M.}~\bibnamefont
  {Ueda}},\ }\href {\doibase 10.1103/PhysRevE.84.051122} {\bibfield  {journal}
  {\bibinfo  {journal} {Phys. Rev. E}\ }\textbf {\bibinfo {volume} {84}},\
  \bibinfo {pages} {051122} (\bibinfo {year} {2011})}\BibitemShut {NoStop}%
\bibitem [{\citenamefont {Holevo}(1998)}]{Holevo1998}%
  \BibitemOpen
  \bibfield  {author} {\bibinfo {author} {\bibfnamefont {A.~S.}\ \bibnamefont
  {Holevo}},\ }\href
  {http://www.citebase.org/abstract?id=oai:arXiv.org:quant-ph/9611023}
  {\bibfield  {journal} {\bibinfo  {journal} {IEEE Transactions on Information
  Theory}\ }\textbf {\bibinfo {volume} {44}} (\bibinfo {year}
  {1998})}\BibitemShut {NoStop}%
\bibitem [{\citenamefont {Schumacher}\ and\ \citenamefont
  {Westmoreland}(1997)}]{Schumacher1997}%
  \BibitemOpen
  \bibfield  {author} {\bibinfo {author} {\bibfnamefont {B.}~\bibnamefont
  {Schumacher}}\ and\ \bibinfo {author} {\bibfnamefont {M.~D.}\ \bibnamefont
  {Westmoreland}},\ }\href {\doibase 10.1103/PhysRevA.56.131} {\bibfield
  {journal} {\bibinfo  {journal} {Phys. Rev. A}\ }\textbf {\bibinfo {volume}
  {56}},\ \bibinfo {pages} {131} (\bibinfo {year} {1997})}\BibitemShut
  {NoStop}%
\bibitem [{\citenamefont {Wilde}(2013)}]{Wilde2013}%
  \BibitemOpen
  \bibfield  {author} {\bibinfo {author} {\bibfnamefont {M.~M.}\ \bibnamefont
  {Wilde}},\ }\href {http://dx.doi.org/10.1017/CBO9781139525343.004} {\emph
  {\bibinfo {title} {Quantum Information Theory}}}\ (\bibinfo  {publisher}
  {Cambridge University Press},\ \bibinfo {year} {2013})\BibitemShut {NoStop}%
\bibitem [{Note2()}]{Note2}%
  \BibitemOpen
  \bibinfo {note} {Note that ${\protect \cal M}$ is not generally an integer
  number.}\BibitemShut {Stop}%
\bibitem [{Note3()}]{Note3}%
  \BibitemOpen
  \bibinfo {note} {In \cite {Schumacher1997} the codes are constructed by
  choosing a number of codewords independently, according to the {\protect \em
  a priori} string probability for each codeword. The choice is supposed to be
  random and the results concerning the probability of error are averaged over
  the different possibilities.}\BibitemShut {Stop}%
\end{thebibliography}%
\end{document}